\documentclass[onecolumn,11pt,letterpaper]{usetex-onecol}

\usepackage{algorithm}
\usepackage{algpseudocode}
\usepackage{booktabs}
\usepackage{graphicx}
\usepackage{url}
\usepackage{color}
\usepackage{multirow}
\usepackage{amsmath}
\usepackage{amsthm}
\usepackage{xspace}
\usepackage{array}
\usepackage{enumitem}
\usepackage{epstopdf}
\usepackage[normalem]{ulem}
\usepackage{microtype}
\usepackage{subcaption}
\usepackage{multicol}
\usepackage{makecell}
\DisableLigatures{encoding=*, family=*}

\usepackage{times}
\usepackage{breakurl}
\usepackage{color}
\usepackage{enumitem}
\usepackage{graphicx}
\usepackage{xspace}
\usepackage{epstopdf}
\usepackage{url}
\usepackage{cite}
\usepackage[colorlinks=true,linkcolor=black,citecolor=red,pdfborder={0 0 0}]{hyperref}
\usepackage{breakurl}
\usepackage{enumitem}
\usepackage{microtype}

\setlist{nolistsep}
\usepackage{mathptmx}
\DeclareMathAlphabet{\mathcal}{OMS}{cmsy}{m}{n}
\DeclareMathAlphabet{\mathrm}{OT1}{bch}{m}{n}
\DeclareMathAlphabet{\mathit}{OT1}{bch}{m}{it}

\newcommand{\para}[1]{{\smallskip\noindent\bf #1}}
\newcommand{\sysname}{ZapRAID\xspace}

\begin{document}

\title{The Design and Implementation of a High-Performance Log-Structured RAID System for ZNS SSDs
\thanks{An earlier version of this article appeared in \cite{wang23}.  In this
extended version, we enhance \sysname to simultaneously achieve intra-zone
and inter-zone parallelism through a careful combination of Zone Append and
Zone Write.  We also include new experiments, including microbenchmarks,
trace-driven experiments, and real-application experiments, to study more
performance aspects of \sysname.}}

\author{Jinhong Li$^1$, Yiyang Geng$^1$, Qiuping Wang$^1$, Shujie Han$^2$,
Patrick P. C. Lee$^1$\\
$^1$The Chinese University of Hong Kong \ \ $^2$Peking University}

\maketitle

\begin{sloppypar}
\begin{abstract} 
Zoned Namespace (ZNS) defines a new abstraction for host software to flexibly
manage storage in flash-based SSDs as append-only zones.  It also provides a
Zone Append primitive to further boost the write performance of ZNS SSDs by
exploiting intra-zone parallelism.  However, making Zone Append effective for
reliable and scalable storage, in the form of a RAID array of multiple ZNS
SSDs, is non-trivial, since Zone Append offloads address management to ZNS
SSDs and requires hosts to specifically manage RAID stripes across multiple
drives.  We propose \sysname, a high-performance log-structured RAID system
for ZNS SSDs by carefully exploiting Zone Append to achieve high write
parallelism and lightweight stripe management.  \sysname adopts a group-based
data layout with a coarse-grained ordering across multiple groups of stripes,
such that it can use small-size metadata for stripe management on a per-group
basis under Zone Append.  It further adopts hybrid data management to
simultaneously achieve intra-zone and inter-zone parallelism through a careful
combination of both Zone Write and Zone Append primitives.  We implement
\sysname as a user-space block device, and evaluate \sysname using
microbenchmarks, trace-driven experiments, and real-application experiments.
Our evaluation results show that \sysname achieves high write throughput and
maintains high performance in normal reads, degraded reads, crash recovery,
and full-drive recovery.
\end{abstract}

\section{Introduction}
\label{sec:intro}

Flash-based solid-state drives (SSDs) have been widely deployed as they have
better performance and reliability than hard-disk drives (HDDs).  Conventional
SSDs expose a block interface, implemented inside the flash translation layer
(FTL), to a host for accessing flash storage. However, such a block interface
also introduces costly internal flash management operations (e.g., device-level
garbage collection and address translation) that lead to unpredictable I/O
performance \cite{hao16} and reduced flash endurance \cite{bjorling21}.  The
recently proposed NVMe Zoned Namespace (ZNS) interface
\cite{zonedstorage,bjorling21} abstracts flash-based SSDs as append-only {\em
zones}, so that the FTL and its costly operations are eliminated, and the
storage management is now shifted to the host.  Compared with conventional
SSDs, ZNS SSDs are shown to achieve higher write throughput, lower tail read
latencies, and less device-level DRAM usage \cite{bjorling21}. 

ZNS SSDs support two write primitives, namely {\em Zone Write} and {\em
Zone Append}, where Zone Append provides opportunities for further performance
gains by exploiting intra-zone parallelism.  Specifically, a ZNS SSD tracks
per-zone {\em write pointers}, such that any write to a zone must specify
the same offset indicated by the write pointer.  Zone Write requires the host
to specify the block address when writing a block, as in conventional SSDs.
This provides backward compatibility with existing applications.  However, in
order to match the offset of a Zone Write with the on-device write pointer of
a zone, the host can only issue one request to each zone at a time, thereby 
limiting intra-zone parallelism.  In contrast, Zone Append eliminates the need
for the host to specify block addresses in writes by fully offloading 
address management to ZNS SSDs.  In Zone Append, the host only specifies the
zone to which a write is issued, and the ZNS SSD returns the address to the
host upon write completion.  Thus, the host can issue multiple Zone Append
commands within one zone to exploit the intra-zone parallelism for improved
write performance. 

Despite the performance gains of Zone Append, its offloading of address
management to ZNS SSDs implies that the host not only cannot directly specify 
the block addresses of writes in a zone, but it also cannot control the
ordering of writes in concurrent Zone Append commands.  This creates new
challenges to storage system designs that require host-level address
management, particularly when applying Redundant Array of Independent Disks
(RAID) \cite{patterson88} to form an array of multiple ZNS SSDs for reliable
and scalable storage. Specifically, in a traditional RAID array, a RAID
controller runs atop multiple drives and organizes data in {\em stripes}, each
of which encodes a set of data blocks into parity blocks and distributes the
data and parity blocks across drives.  For efficient data repair, the RAID
controller statically assigns the same block address in each drive for the
blocks of the same stripe, so that the repair of any lost block can directly
retrieve the available data and parity blocks of the same stripe from other
drives for decoding.  However, under Zone Append, it is infeasible for the
RAID controller to manage block addresses in a zone; instead, the RAID
controller needs to maintain dedicated address mapping information to specify
the block locations for each stripe, which unavoidably incurs performance
penalties to stripe management.


We present \sysname, a high-performance RAID system for ZNS SSDs by carefully
exploiting Zone Append to achieve high write performance via intra-zone
parallelism; in the meantime, \sysname supports lightweight stripe management
in terms of: (i) querying stripe metadata during degraded reads, crash
recovery, and full-drive recovery, (ii) memory usage for indexing, and (iii)
persistent storage for metadata.  
\sysname builds on log-structured RAID (Log-RAID)
\cite{chiueh14, colgrove15, ioannou18, kim19}, which issues sequential writes
to a RAID array, and extends Log-RAID with a novel {\em group-based data
layout}.  Specifically, \sysname partitions stripes into stripe groups and
issues Zone Append to the stripes within the same stripe group for high write
performance.  The group-based data layout organizes stripes with a
coarse-grained ordering and enables \sysname to manage stripes efficiently on a
per-group basis.  

From our measurement study (\S\ref{subsec:vs}),
we observe that Zone Append effectively exploits intra-zone parallelism when a
workload is dominated by small writes and when writes are issued to a single
zone of a ZNS SSD at a time, while Zone Write is beneficial when a workload is
dominated by large writes or when writes are concurrently issued to multiple
zones of a ZNS SSD (in which Zone Write exploits better inter-zone parallelism
for large writes).  To this end, \sysname adopts {\em hybrid data management}
to manage small and large writes across multiple zones in a ZNS SSD, so as to
simultaneously achieve intra-zone and inter-zone parallelism through a careful
combination of both Zone Append and Zone Write primitives. 

In summary, we make the following contributions:
\begin{itemize}[leftmargin=*]
\item
We conduct a measurement study to carefully examine the trade-off between Zone
Append and Zone Write on a ZNS SSD (\S\ref{subsec:vs}). 
\item
We design \sysname, a high-performance Log-RAID system for ZNS SSDs. \sysname
builds on two core ideas: (i) a group-based data layout under Zone Append
(\S\ref{subsec:group}) and (ii) hybrid data management (\S\ref{subsec:hybrid})
under a combination of Zone Write and Zone Append. \sysname further addresses
reliability in terms of metadata persistence and crash consistency
(\S\ref{subsec:crash}). 
\item
We prototype \sysname as a user-space block device with the Storage
Performance Development Kit (SPDK) \cite{spdk}, such that \sysname exports a
block interface for compatibility with general applications and can
parallelize its internal operations with multiple SPDK threads. 
\item
We evaluate our \sysname prototype on real ZNS SSD devices using
microbenchmarks, trace-driven experiments, and real-application experiments.
When writes are issued to a single zone per ZNS SSD, \sysname increases the
write throughput by up to 77.2\% and reduces the tail latency by up to
36.4\% compared with the exclusive use of Zone Write, and also scales to high
write throughput under FEMU emulation \cite{li18}.  When writes are issued to
multiple zones per ZNS SSD, \sysname always maintains the highest throughput
and lowest tail latency compared with the exclusive use of either Zone Write
or Zone Append.  Furthermore, \sysname maintains high performance in normal
reads, degraded reads, crash recovery, and full-drive recovery. 
\end{itemize}

We now make the source code available on
\textbf{https://github.com/adslabcuhk/zapraid}. 

\section{Background and Motivation}
\label{sec:background}

We provide the basics of ZNS (\S\ref{subsec:zns}) and study the trade-off
between Zone Write and Zone Append (\S\ref{subsec:vs}). We also review the
background of Log-RAID and state the challenges of deploying Log-RAID on ZNS
SSDs (\S\ref{subsec:lograid}). 

\subsection{Zoned Namespace (ZNS)}
\label{subsec:zns}

\begin{figure}[!t]
\centering
\includegraphics[width=5.0in]{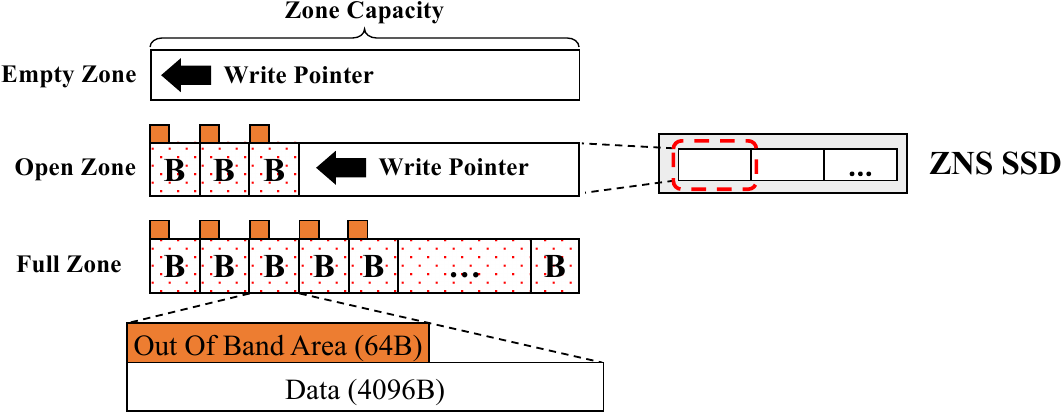}
\vspace{-3pt}
\caption{Architecture of a ZNS SSD based on the Western Digital Ultrastar DC
ZN540 model \cite{zn540}.}
\label{fig:zns}
\end{figure}

The Zoned Namespace (ZNS) interface \cite{zonedstorage,bjorling21} is a new
NVMe Command Set \cite{nvme} that exposes flash-based SSD internals to the
host for host-level storage management.  Figure~\ref{fig:zns} shows the
architecture of a ZNS SSD based on the Western Digital Ultrastar DC ZN540
model, which we use in this paper and also in previous studies (e.g., 
\cite{bergman22, kim23, lee23, seo23, doekemeijer23}).  
The ZNS interface abstracts a ZNS SSD as
append-only {\em zones}, each of which has a maximum size called the 
{\em zone capacity}.  Each zone organizes data in logical blocks (or {\em
blocks} in short in this paper), and each block is mapped to a physical flash
page.  Each flash page also has an out-of-band area for metadata storage
(of size tens of bytes) located alongside the page content. Note that the
out-of-band area is also available in general NVMe SSDs, and its space is not
included in the total zone capacity.  In this paper, we assume that the block
size is 4\,KiB and the out-of-band area size is 64~bytes, both of which can be
configured in the ZN540 model when a ZNS SSD is formatted. 

Within a zone, the blocks can be randomly read, but must be sequentially
written; meanwhile, multiple zones can be read/written concurrently.
Specifically, a ZNS SSD maintains a per-zone {\em write pointer} that
specifies the offset of the next write.  It mandates that any new write
to a zone must be issued to the offset referenced by the write pointer.  Upon
the completion of each write request, the write pointer is incremented by the
number of written blocks. 

Each zone in a ZNS SSD is associated with a {\em state}. An {\em empty} zone
means that the zone has not been written any block. Once a block is written
to an empty zone, the zone becomes an {\em open} zone and can accept new
writes before it is full.  Typically, the number of open zones in a ZNS SSD is
bounded.  Once the number of blocks written to a zone reaches the zone
capacity, the zone becomes a {\em full} zone. The host can also explicitly
{\em finish} an open zone to turn it into a full zone. To overwrite the data
in a zone, the host must first explicitly {\em reset} the zone to erase any
data in a block and rewind the write pointer to the beginning of the zone, and
the zone becomes an empty zone after reset.

\subsection{Zone Write versus Zone Append}
\label{subsec:vs}

Recall from \S\ref{sec:intro} that ZNS provides the Zone Write and Zone Append
primitives for a host to issue writes to a zone.  Their key difference is that
Zone Write requires the host to issue writes serially and specify the offset
in each write, while Zone Append offloads address assignment to ZNS SSDs and
enables the host to issue concurrent writes that exploit intra-zone
parallelism.  Existing ZNS devices and host software typically allow only one
outstanding Zone Write per zone in order to maintain the write pointer
invariant; otherwise, concurrent Zone Write and Zone Append commands may
violate the sequential-write requirement.

\begin{figure}[!t]
\centering
\begin{tabular}{ccc}
\multicolumn{3}{c}{
\includegraphics[width=2in]{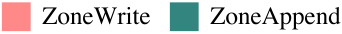}
}\\
\includegraphics[width=2in]{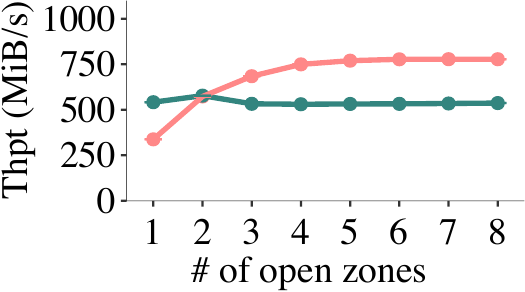} & 
\includegraphics[width=2in]{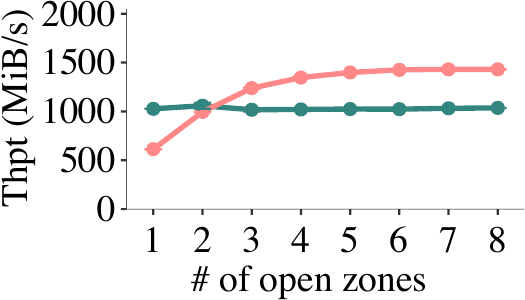} &
\includegraphics[width=2in]{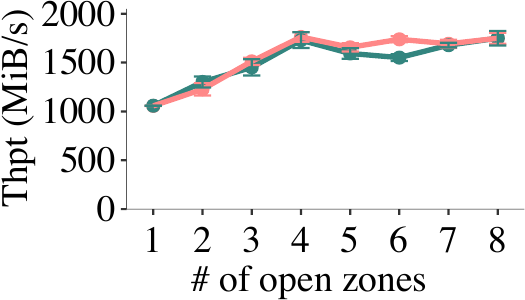} \\
{\small (a) 4\,KiB requests} & 
{\small (b) 8\,KiB requests} &
{\small (c) 16\,KiB requests} 
\end{tabular}
\vspace{-9pt}
\caption{Write throughput of Zone Write and Zone Append versus the number of
open zones under various request sizes.}
\label{fig:motivation}
\end{figure}

We conduct evaluation on a single ZN540 ZNS SSD \cite{zn540} in our testbed (see
\S\ref{subsec:settings} for testbed details) to show how Zone Append improves
write performance over Zone Write.  We issue write requests of 4\,KiB, 8\,KiB,
and 16\,KiB to the ZNS SSD.  We also vary the number of open zones in the
ZNS SSD.  We write a total of 512\,GiB of data and measure the write
throughput.  We report the average throughput results over five runs, with the
error bars showing the 95\% confidence intervals under the Student's
t-distribution (note that some error bars may be invisible due to small
deviations).  Figure~\ref{fig:motivation} shows the write throughput results.  

\para{Writes to a single open zone.}
We first consider the case when the number of
open zones is one.  In this case, we issue writes to a single zone at a time;
if the zone is full, we issue writes to a different zone.  Zone Write achieves
a write throughput of 337.6\,MiB/s, 613.6\,MiB/s, and 1,050.0\,MiB/s for
4\,KiB, 8\,KiB, and 16\,KiB requests, respectively. Recall that a ZNS SSD can
have only one outstanding Zone Write at any time.  We find that the write
throughput of Zone Write cannot further increase even if we issue multiple
concurrent write requests of Zone Write.  In contrast, Zone Append achieves
a write throughput of 541.5\,MiB/s, 1,026.6\,MiB/s, and 1,050.1\,MiB/s for
4\,KiB, 8\,KiB, and 16\,KiB requests, respectively, with four concurrent write
requests of Zone Append\footnote{The ZN540 specification \cite{zn540} states
that the write throughput can reach 2,000\,MiB/s. This number corresponds to
the best-case performance and is obtained when large requests (128\,KiB) are
issued to all open zones.}.  This shows that Zone Append can increase the
write throughput via intra-zone parallelism.

Note that in this single-zone experiment, even if we issue more than four
concurrent write requests, the write throughput of Zone Append for 4\,KiB and
8\,KiB requests does not further increase, as it already saturates intra-zone
parallelism for this specific ZN540 model. Also, Zone Write already achieves
the maximum write throughput of a zone in ZN540 for 16\,KiB requests, so Zone
Append does not see more growth.  

\para{Writes to multiple open zones.}
We further consider the case when the number of open zones is larger than one,
so as to examine how {\em inter-zone} parallelism affects the effectiveness
of Zone Write and Zone Append.  We create multiple open zones and issue writes
to the open zones in parallel.  We set the number of concurrent write requests
for each open zone as one and four for Zone Write and Zone Append,
respectively, based on the previous results in the single-zone experiment.
From Figure~\ref{fig:motivation}, we observe that the write throughput of Zone
Write is generally higher when there are more open zones (e.g., its write
throughput reaches 777.1\,MiB/s under six open zones for 4\,KiB write requests
(Figure~\ref{fig:motivation}(a))).  On the other hand, Zone Append can only
achieve a maximum write throughput of 577.5\,MiB/s under two open zones for
4\,KiB requests (Figure~\ref{fig:motivation}(a)). A possible reason is that
the current firmware implementation of Zone Append is more computationally
intensive when writes are issued to more open zones in the ZN540 model, so
Zone Append has even lower write throughput than Zone Write under a larger
number of open zones due to the limited computational power in existing
hardware.  We conjecture that this limitation could be addressed in future ZNS
SSD models.  When the request size is 8\,KiB (Figure~\ref{fig:motivation}(b)),
Zone Write can scale to 1,430.7\,MiB/s under eight open zones, while Zone
Append can only achieve a maximum write throughput of 1,058.6\,MiB/s under two
open zones.  When the request size is 16\,KiB
(Figure~\ref{fig:motivation}(c)), Zone Write and Zone Append have very small
performance differences (e.g., up to around 1,750\,MiB/s under eight open
zones), implying large writes can effectively exploit inter-zone parallelism
in both Zone Write and Zone Append. 

\para{Main observations.}  We observe that for small writes (e.g., of
size 4\,KiB or 8\,KiB), Zone Append achieves higher write throughput than Zone
Write in the single-zone setting by exploiting intra-zone parallelism, while
Zone Write scales better and achieves higher write throughput than Zone Append
when the number of open zones increases by better exploiting inter-zone
parallelism.  For large writes (e.g., of size 16\,KiB), both Zone Write and
Zone Append have similar write throughput for different numbers of open zones.   
A RAID system based on ZNS SSDs should adequately combine Zone Append and Zone
Write to accommodate the scenarios with different numbers of open zones.  Even
though Zone Append only shows high performance when write requests are issued
to one or a few open zones, it is still beneficial for the
following scenarios. First, shared storage systems need performance isolation
\cite{chang15,huang17,kim18}.  Specifically, since a zone is typically mapped
to only a subset of physical flash channels or chips \cite{bae22}, a shared
storage system based on ZNS SSDs should allocate only a limited number of open
zones to applications that do not have high write bandwidth demands.  This
isolates the write traffic of the applications to a subset of flash channels
and limits their impact on other co-located applications. Second, it
is common to have mixed request sizes in many real-world workloads.  For
example, practical cloud block storage workloads have 60\% of write requests
that are at most 4\,KiB and 25\% of write requests that are at least 16\,KiB
\cite{li20}.  By separating small and large writes into different open zones
(\S\ref{subsec:hybrid}), we can use Zone Append and Zone Write for different
groups of writes to improve the overall performance (\S\ref{subsec:micro}).
Thus, Zone Append can help such applications achieve high write performance
through intra-zone parallelism and sustain the bursts of a large number of
write requests if necessary, even with only one or a few open zones
per ZNS SSD being used. 

\subsection{Log-structured RAID (Log-RAID)}
\label{subsec:lograid}

RAID \cite{patterson88} forms an array of storage drives to improve capacity,
performance, and fault tolerance in storage.  We focus on parity-based RAID,
which organizes the RAID storage space in units of {\em stripes}.  Each stripe
contains multiple {\em chunks} that span across all drives (one chunk per
drive), where each chunk contains a fixed number of contiguous blocks.  A
$(k+m)$-RAID array forms a stripe with $k+m$ chunks by encoding a set of $k$
data chunks into $m$ parity chunks, such that any $k$ out of the $k+m$ chunks
of the same stripe can decode the $k$ data chunks.  It distributes the $k+m$
chunks of each stripe across $k+m$ drives, so as to provide fault tolerance
against any $m$ failed drives.  To make data repair efficient, RAID performs
{\em static mapping}, which aligns each data or parity chunk of the same
stripe at the same offset of a drive.  Thus, when recovering any lost chunk,
the RAID controller can deterministically retrieve $k$ available chunks of the
same stripe at the same offset from other drives for decoding. 

Traditional RAID is designed for HDDs \cite{patterson88}, which adopt in-place
updates to overwrite blocks.  In contrast, SSDs adopt out-of-place updates.
Small random writes to SSDs are shown to trigger frequent device-level garbage
collection that degrades both I/O performance and flash endurance
\cite{chen09,kim08,min12}.  Parity-based RAID is even more vulnerable to small
random writes, which trigger parity updates that further aggravate performance
and endurance degradations.  Thus, some studies \cite{chiueh14, colgrove15,
ioannou18, kim19} propose Log-RAID, which applies the log-structured file
system design \cite{rosenblum92} to SSD RAID by issuing sequential host-level
writes.

Log-RAID manages stripes in append-only {\em segments}.  Each segment holds
a number of stripes (up to some pre-specified capacity) and is mapped to $k+m$
fixed-size contiguous areas that reside in $k+m$ drives (one contiguous area
per drive), so as to provide fault tolerance against any $m$ failed drives.  A
segment is {\em open} if it does not reach its capacity and hence can accept
new writes.  Log-RAID aggregates newly written chunks as new stripes.  It
writes the new $k+m$ chunks of each stripe to the same offset of the $k+m$
contiguous areas in an append-only manner, such that $k+m$ chunks of the same
stripe are aligned at the same offset of the $k+m$ contiguous areas (i.e.,
static mapping).  If an open segment reaches its full capacity, Log-RAID seals
it into a {\em sealed} segment, and creates a new open segment from the free
contiguous areas in the underlying drives. 

By writing blocks as new stripes in an append-only manner, Log-RAID needs to
perform garbage collection to reclaim the space from stale blocks.  When
garbage collection is triggered (say, when the available space drops below some
threshold), Log-RAID selects a sealed segment by some policy (e.g., using a
greedy algorithm to select the one with the largest fraction of stale blocks),
rewrites all non-stale blocks into a new open segment, and releases the space
of the selected sealed segment for reuse.

Unlike traditional RAID, Log-RAID necessitates dedicated in-memory data
structures for stripe management.  Specifically, it keeps a {\em segment
table} to map each segment to its corresponding $k+m$ contiguous areas.  It
also keeps an {\em address table} to map each block address (identified by the
application) to the corresponding segment, drive, and offset within the
contiguous areas.  To repair a lost block, Log-RAID first identifies the
segment and drive in which the lost block resides from the address table.  It
then identifies the contiguous areas of the other drives for the segment from
the segment table.  It retrieves $k$ available chunks at the same offset in
other contiguous areas to decode the lost chunk (as in the static mapping in
traditional RAID), and finally recovers the lost block. 

Log-RAID is a natural fit for ZNS SSDs by mapping each contiguous area of a
drive to a zone, which organizes writes in an append-only manner.  However,
applying Zone Append in Log-RAID is non-trivial. Since Zone Append offloads
address management to ZNS SSDs, the chunks of the same stripe may now be
written to different offsets within the zones.  To repair any lost block, we
can no longer directly retrieve the available blocks from the same offset of
other zones based on static mapping; instead, we need to keep track of the
offsets of the chunks of the same stripe.  How to efficiently manage chunk
locations when Zone Append is used in Log-RAID for ZNS SSDs is an important
design issue. 

\section{\sysname Design}
\label{sec:design}

\sysname is an extended Log-RAID design for ZNS SSD arrays with three key
design goals: (i) {\em high performance}, i.e., \sysname achieves high
throughput and low latency via a combination of Zone Append and Zone Write;
(ii) {\em lightweight stripe management}, i.e., \sysname keeps both low
memory usage for index structures and low storage overhead for metadata
persistence, while supporting fast metadata queries for locating the
chunks of a stripe; and (iii) {\em reliability}, i.e., \sysname protects data
and metadata against failures. 

\sysname achieves the above goals with two main designs.  The first design is
a {\em group-based data layout}.  Specifically, when \sysname uses Zone
Append for intra-zone parallelism, it maintains a coarse-grained ordering of
stripes by organizing stripes in {\em stripe groups}, such that it issues Zone
Append for the stripes on a per-group basis, so as to efficiently manage
stripes for each stripe group instead of in a system-wide manner. It further
enforces reliability by maintaining consistency for stripes and index
structures during crash recovery. The second design is {\em hybrid data
management}.  \sysname serves the write requests of small and large sizes
using different segments, such that a segment is configured with a small
(resp.  large) chunk size to serve small (resp. large) writes.  This allows
\sysname to serve small writes with low latencies and maintain high overall
write throughput.  Also, by managing segments in a hybrid manner, \sysname can
further apply Zone Append and Zone Write for different segments to achieve
intra-zone and inter-zone parallelism, respectively.

\subsection{Design Overview}
\label{subsec:overview}

\begin{figure}[!t]
\centering
\includegraphics[width=5.0in]{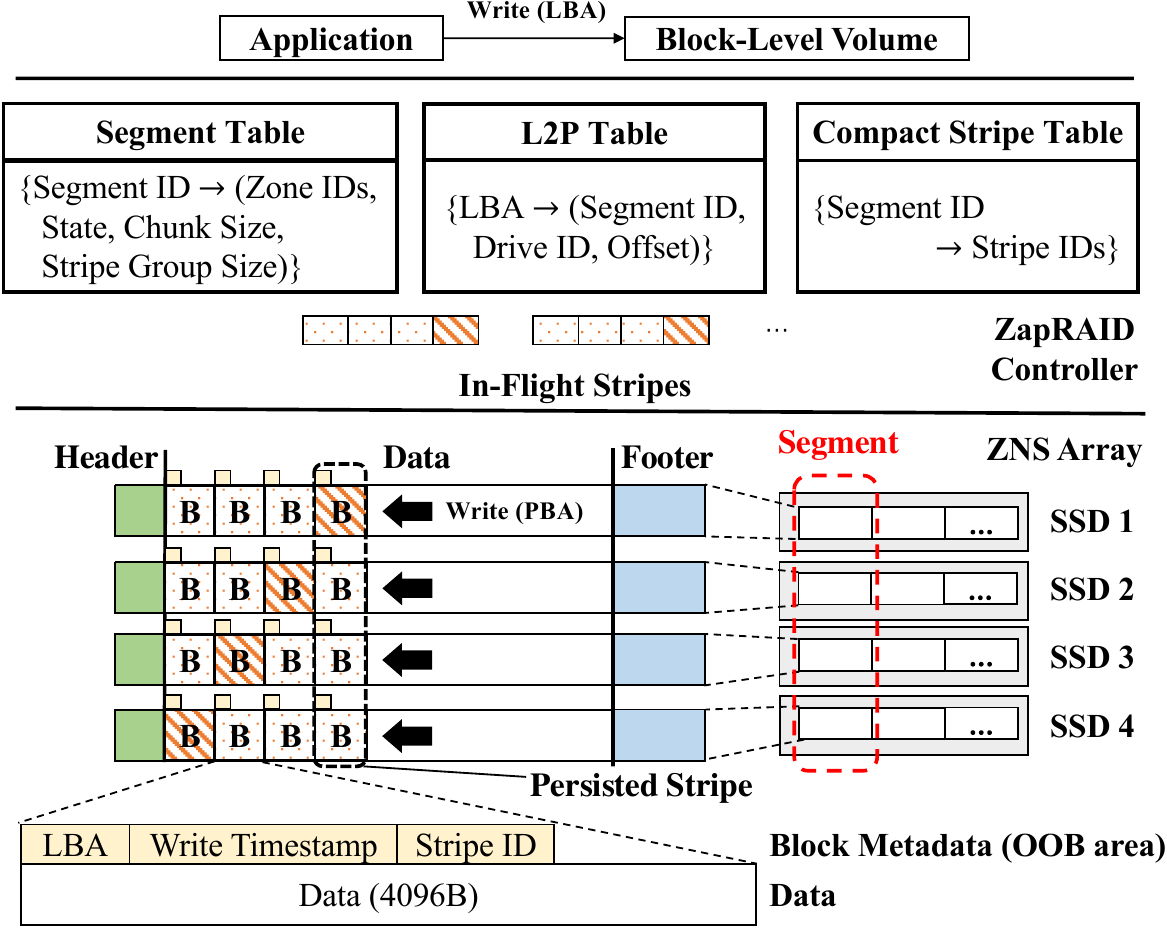}
\vspace{-3pt}
\caption{\sysname architecture. Here, we show a (3+1)-RAID-5 array and the
layout of a segment of four zones, where the parity chunks are rotated across
the drives.}
\label{fig:zapraid}
\vspace{-6pt}
\end{figure}

{\bf Architecture.}  We first introduce the architecture of \sysname and show
how it augments Log-RAID on an array of multiple ZNS SSDs, as shown in
Figure~\ref{fig:zapraid}.  \sysname exposes a block-level {\em volume} that
supports random reads/writes.  We assume that the block size is 4\,KiB. A user
application can issue reads/writes of an arbitrary number of blocks, each of
which is identified by a logical block address (LBA).  Each drive corresponds
to a ZNS SSD, and \sysname maps each contiguous area of a drive under Log-RAID
to a zone, which only allows sequential writes.  Let $Z$ be the total number
of zones in a drive, so there are $Z$ segments in a ZNS SSD array.  For
example, a 4\,TiB ZN540 ZNS SSD \cite{zn540} configures $Z=$~3,690 zones.  Each
of the drives, segments, zones, and stripes is associated with an identifier
(ID). 

\para{Segment organization.}  Each segment corresponds to $k+m$ zones in
$k+m$ drives.  In \sysname, each segment comprises three regions that span
across $k+m$ drives: the {\em header region} and the {\em footer region} for
keeping metadata for crash consistency (\S\ref{subsec:crash}), and the {\em
data region} for storing data and parity chunks.  Specifically, the header
region stores the RAID scheme (including RAID-5/RAID-6, $k$, and $m$)
and the segment-specific information (including the zone IDs of all zones from
the $k+m$ drives in the segment, the chunk size, and the stripe group size
(defined in \S\ref{subsec:group})).  The footer region keeps the {\em block
metadata}, including the LBA, write timestamp, and stripe ID of each block in
the data region.

All three regions have pre-specified sizes. The header region contains exactly
one stripe of $k+m$ chunks, each of which resides at the beginning of a zone.
The data region contains a fixed number of stripes, denoted by $S$, following
the header region. For each segment, each chunk contains a fixed number of
blocks, denoted by $C$, so the number of blocks in the data region is $S\cdot
C$.  Suppose that the LBA size is 8~bytes, the write timestamp size is
8~bytes, the stripe ID size is 4~bytes, and the block size is 4\,KiB.  Thus,
each block in the footer region can store the block metadata of
$\lfloor\tfrac{4096}{20}\rfloor = 204$~blocks, so the footer region occupies
$\lceil\tfrac{S\cdot C}{204}\rceil$~blocks following the data region.  For
example, the capacity of a zone in a ZN540 ZNS SSD \cite{zn540} is
1,077\,MiB (or equivalently, 275,712 blocks).  If the chunk size is 4\,KiB
(i.e., $C=1$), then the header, data, and footer regions occupy 1~block,
274,366~blocks, and 1,345~blocks in a zone, respectively.

When \sysname creates a new open segment across $k+m$ drives, it first writes
the header region.  It then writes the stripes to the data region.  Depending
on the RAID scheme, the parity chunks of all stripes may reside in a fixed
drive (for RAID-4) or be rotated across the drives (for RAID-5 and RAID-6)
(e.g., see the RAID-5 example in Figure~\ref{fig:zapraid}).  When the data
region reaches its pre-specified size, \sysname writes the block metadata
of all blocks in the segment to the footer region.  Finally, it seals the
segment.

\para{Block metadata.} \sysname stores the block metadata for each block
in the out-of-band area of the corresponding flash page (\S\ref{subsec:zns})
for persistence.
Each block in a data chunk has its LBA, write timestamp, and stripe ID as its
block metadata.  \sysname provides fault tolerance for the block metadata: for
LBAs and write timestamps, \sysname generates parity-based redundancy for them
from all blocks of the data chunks in the same stripe and stores the parity
results in the block metadata of the blocks of the parity chunks, while for
stripe IDs, \sysname replicates them into all the data and parity chunks in
the same stripe.

Recall that the footer region also stores the block metadata (i.e., the LBA,
write timestamp, and stripe ID) for all blocks in the segment, so \sysname
keeps two copies of block metadata (i.e., in the out-of-band area of each
flash page and the footer region).  We argue that both copies are necessary
and serve different purposes: the block metadata in the out-of-band area 
associated with each block provides persistence for block writes, while the
block metadata in the footer region provides fast crash recovery
(\S\ref{subsec:crash}). 

\para{In-memory items.} \sysname keeps a number of in-memory {\em
in-flight stripes} for newly written blocks.  Each in-flight stripe is kept in
memory until all of its $k$ data chunks and $m$ parity chunks are formed
and persisted.  To maintain durability, \sysname acknowledges the writes of an
in-flight stripe only after the whole in-flight stripe is persisted (note that
acknowledging the write of each block can lead to data loss if a drive storing
an acknowledged block fails, but the parity chunks are not yet generated).  

\sysname also maintains three in-memory index structures: (i) the {\em
segment table}, which maps each segment ID to its corresponding $k+m$
zones (identified by the zone IDs in the respective $k+m$ drives), 
segment state, chunk size, and stripe group size; (ii) the {\em
logical-to-physical (L2P) table}, which maps the LBA of each block issued by an
application to the {\em physical block address (PBA)} (i.e., the segment ID,
the drive ID, and the offset in the respective zone); and (iii) the {\em
compact stripe table}, which maps each segment ID to the stripe IDs of all
chunks in the segment.  Both the segment table and L2P table are similarly
found in Log-RAID.  The compact stripe
table is newly introduced to \sysname since Zone Append can make the chunks of
the same stripe reside at different offsets across the drives
(\S\ref{subsec:lograid}).  We show how the compact stripe table is designed and
how we reduce its size via the group-based data layout in \S\ref{subsec:group}.
Note that \sysname ensures fault tolerance for the index structures by
persisting the segment-to-zones mappings in the segment table into the header
region of each segment and persisting the LBAs, write timestamps, and stripe
IDs as block metadata into the out-of-band area associated with each block.

\para{Offloading L2P table entries to ZNS SSDs.} The L2P table incurs large
memory footprints if it keeps all entries in memory.  For example, our current
implementation supports a RAID array of four 4\,TiB ZN540 ZNS SSDs (i.e., a
total capacity of 16\,TiB) and uses a 4-byte L2P entry for each 4\,KiB block.
Thus, the total memory size of the L2P table is 16\,GiB.
Such large L2P table footprints are common in all Log-RAID designs
\cite{chiueh14,colgrove15,ioannou18,kim19}.  Some Log-RAID designs keep part
of the L2P table entries on disk to mitigate the memory overhead
\cite{chiueh14,colgrove15}.  

\sysname by default keeps the whole L2P table in memory, yet it also supports
the offloading of part of the L2P table entries to ZNS SSDs.  It uses the CLOCK
algorithm \cite{corbato68} to evict the non-recently used L2P entries to SSDs. 
Specifically, it divides the L2P table entries into multiple groups, each of
which has 1,024 contiguous entries that can be fit into a 4\,KiB block
(assuming that each entry is of size four bytes).  It maintains an in-memory
bitmap, in which each bit tracks if the corresponding entry group has been
accessed, and a pointer that references the entry group to be considered. It
sets the tracked bit to one if any entry in the corresponding entry group has
been accessed.  Suppose that the L2P table size reaches the memory size limit.
\sysname then searches for an entry group to evict, starting from the one
referenced by the pointer.  If the tracked bit of an entry group is one,
\sysname resets the tracked bit to zero, moves the pointer to the next entry
group, and continues the search; otherwise, if the tracked bit of the entry
group is zero, \sysname selects the entry group to evict, moves the pointer to
the next entry group, and stops the search. 

\sysname maps the entry group into a 4\,KiB {\em mapping block}. It writes the
mapping block together with other user-written blocks, so as to avoid
consuming additional open zone resources.  It maintains a separate
in-memory mapping table to record the PBA of each on-SSD mapping block. Note
that the size of the in-memory mapping table is much smaller, as each of its
entries can track 1,024 L2P table entries (e.g., in our case, the size of the
in-memory mapping table is 16\,MiB for a RAID array of four 4\,TiB ZN540 ZNS
SSDs).  It also keeps the LBA of the first entry of the mapping block in the
LBA field of the block metadata for crash recovery (\S\ref{subsec:crash}).  

It is important to distinguish between the mapping blocks and user-written
blocks during the crash recovery of the L2P table (\S\ref{subsec:crash}).
Thus, \sysname marks the least significant bit of the LBA field in the block
metadata as one for a mapping block. It is feasible since the LBA field of a
user-written block is currently aligned at multiples of 4\,KiB, and its 12
least significant bits are all zeros.  

\subsection{Group-Based Data Layout}
\label{subsec:group}

\sysname exploits Zone Append for intra-zone parallelism, yet it needs to
track the offsets of the chunks of the same stripe under Zone Append.  Thus,
\sysname adopts a {\em group-based data layout} to organize the stripes of a
segment with a coarse-grained ordering for low stripe management overhead.
Specifically, for a segment where Zone Append is issued, \sysname partitions a
fixed number of contiguous stripes, denoted by $G$, within the segment into 
{\em stripe groups}, where $G$ is a configurable parameter.  For each stripe
group, it first issues Zone Append for all stripes within the same stripe
group, such that all chunks of each stripe must reside in the same stripe
group but may reside in different offsets within the stripe group. Each stripe
group is in the same offset ranges across all zones, so the offset ranges of
its blocks can be identified via static mapping.  Most importantly, \sysname
only needs to track a small number of stripes within each stripe group, so it
can use fewer bits for the metadata information to save significant memory
usage. Note that we treat $G=1$ as a special case in which each stripe group
issues Zone Write instead of Zone Append for all stripes.

\begin{figure}[!t]
\centering
\includegraphics[width=5.0in]{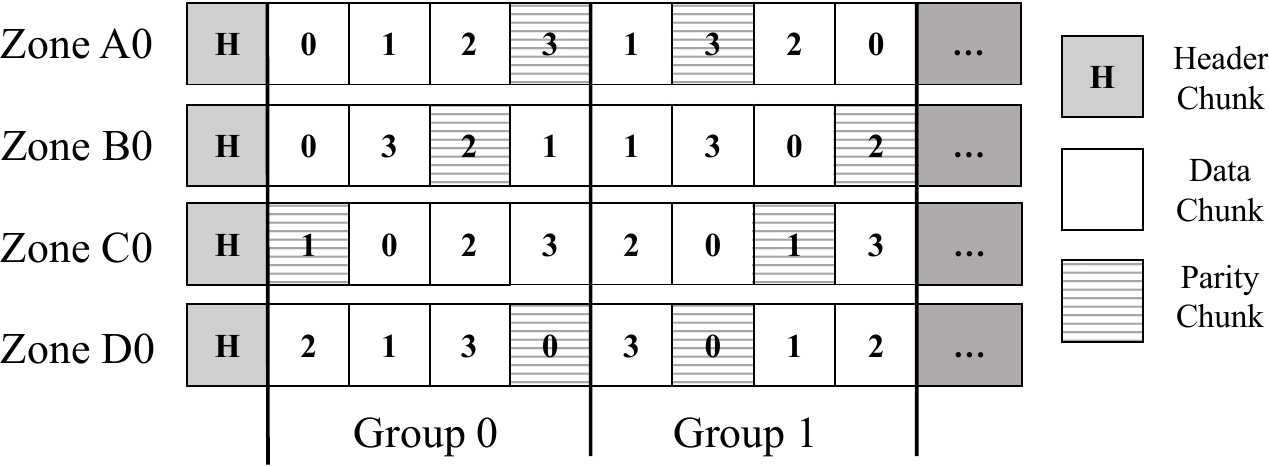}
\vspace{-3pt}
\caption{Group-based data layout of a (3+1)-RAID-5 array, where $G=4$. Each
chunk is labeled with its stripe ID in the group, while the chunks with the
same stripe ID in the same stripe group form a stripe.}
\label{fig:group}
\vspace{-6pt}
\end{figure}

Figure~\ref{fig:group} depicts one segment with $G=4$ stripes per stripe group
in a (3+1)-RAID-5 array.  Within the segment, the data region comprises
a fixed number of stripe groups, each of which further comprises a fixed
number of stripes. In general, the number of stripe groups in a segment is
determined by both $S$ (i.e., the number of stripes in the segment) and $G$.
Each stripe in a stripe group is associated with a unique stripe ID, which can
be viewed as a sequence number of when the stripe is generated.  Due to Zone
Append, the chunks of the same stripe may reside in different offsets, as
shown in Figure~\ref{fig:group}.  

\sysname tracks the stripe IDs of all stripe groups in the compact stripe
table. Specifically, for each segment where Zone Append is issued, the compact
stripe table stores a two-dimensional $(k+m)\times S$ matrix, in which each
entry stores the stripe ID of each chunk in the segment.  In the matrix, the
$i$-th row corresponds to the zone of the $i$-th drive of the RAID array, and
the $j$-th column corresponds to the $j$-th chunk in the segment.  Given the
LBA of a block, \sysname identifies the PBA from the L2P table, calculates the
physical location of the chunk based on the PBA, and retrieves the stripe ID
from the compact stripe table.  

\para{Trade-off analysis.} The choice of $G$ determines the trade-off
between the degree of intra-zone parallelism via Zone Append and the
stripe management overhead.  A larger $G$ allows more stripes to be issued via
Zone Append in parallel, but it also increases the stripe management overhead.
In the extremes, when $G=1$, all stripes are issued via Zone Write; when
$G=S$, there is only one stripe group in each segment (i.e., the group-based
data layout is disabled for Zone Append).

We analyze the stripe management overhead in two aspects: the maximum
memory usage and query overhead of the compact stripe table.  For the maximum
memory usage, each stripe ID is represented in $\lceil \log_{2} G \rceil$
bits.  Thus, the maximum memory size of the compact stripe table is
$(k+m)\cdot S\cdot Z\lceil\log_2 G\rceil$~bits; the memory size
reaches the maximum when all $Z$ segments are written. For the query
overhead, we measure the number of entries in the compact stripe table being
accessed during a degraded read (\S\ref{subsec:all}), which is $k \cdot G$.
With a proper choice of $G$, \sysname can achieve high write performance via
Zone Append, while limiting the memory usage and query overhead of the compact
stripe table.  For example, we consider a (3+1)-RAID-5 array of four 4\,TiB
ZN540 ZNS SSDs and a chunk size of 4\,KiB, where $S=$~274,160 and $Z=$~3,690
(\S\ref{subsec:overview}).  We set $G=256$, our default setting in the
implementation.  The maximum memory size of the compact stripe table is
3.77\,GiB (i.e., one byte per stripe ID); the query overhead is to
access 768 entries, which translates to only around 1$\mu$s from our
evaluation.  

In contrast, the case of $G=S$ (i.e., without the group-based data
layout) translates to 19~bits per stripe ID and hence 8.95\,GiB of memory for
the compact stripe table.  Also, its high query overhead significantly
increases the degraded read latency (\S\ref{subsec:micro}). 

Our prototype rounds up the length of a stripe ID to the nearest byte and
represents a stripe ID in $\lceil \lceil \log_2 G \rceil/8 \rceil$~bytes.
Representing stripe IDs in units of bytes, instead of bits, simplifies data
alignment without compromising our conclusion. Thus, for $G=S$, our
prototype consumes 3~bytes per stripe ID, leading to 11.3\,GiB of memory for
the compact stripe table. 

\subsection{Hybrid Data Management}
\label{subsec:hybrid}

If an application can write data to multiple open zones in a ZNS SSD,
\sysname can create multiple open segments across the open zones and combine
Zone Append and Zone Write to achieve both intra-zone and inter-zone
parallelism.  Recall that Zone Write achieves higher write throughput than Zone
Append under multiple open zones (i.e., higher inter-zone parallelism) and
maintains high write throughput for large writes.  Unlike Zone Append, Zone
Write can maintain static mappings in Log-RAID and hence does not require the
compact stripe table for stripe management.  Thus, the core idea of \sysname
is to reserve an open segment for small writes to exploit intra-zone
parallelism based on Zone Append, while favoring Zone Write for large writes
to exploit inter-zone parallelism. 

To this end, \sysname adopts {\em hybrid data management}.  It classifies the 
open segments into two types, namely {\em small-chunk segments} and {\em
large-chunk segments}, to serve small and large writes, respectively.
It applies Zone Append to one of the small-chunk segments based on the
group-based data layout (\S\ref{subsec:group}), and applies Zone Write to the
remaining small-chunk segments and all large-chunk segments. 

\begin{figure}[!t]
\centering
\includegraphics[width=6.4in]{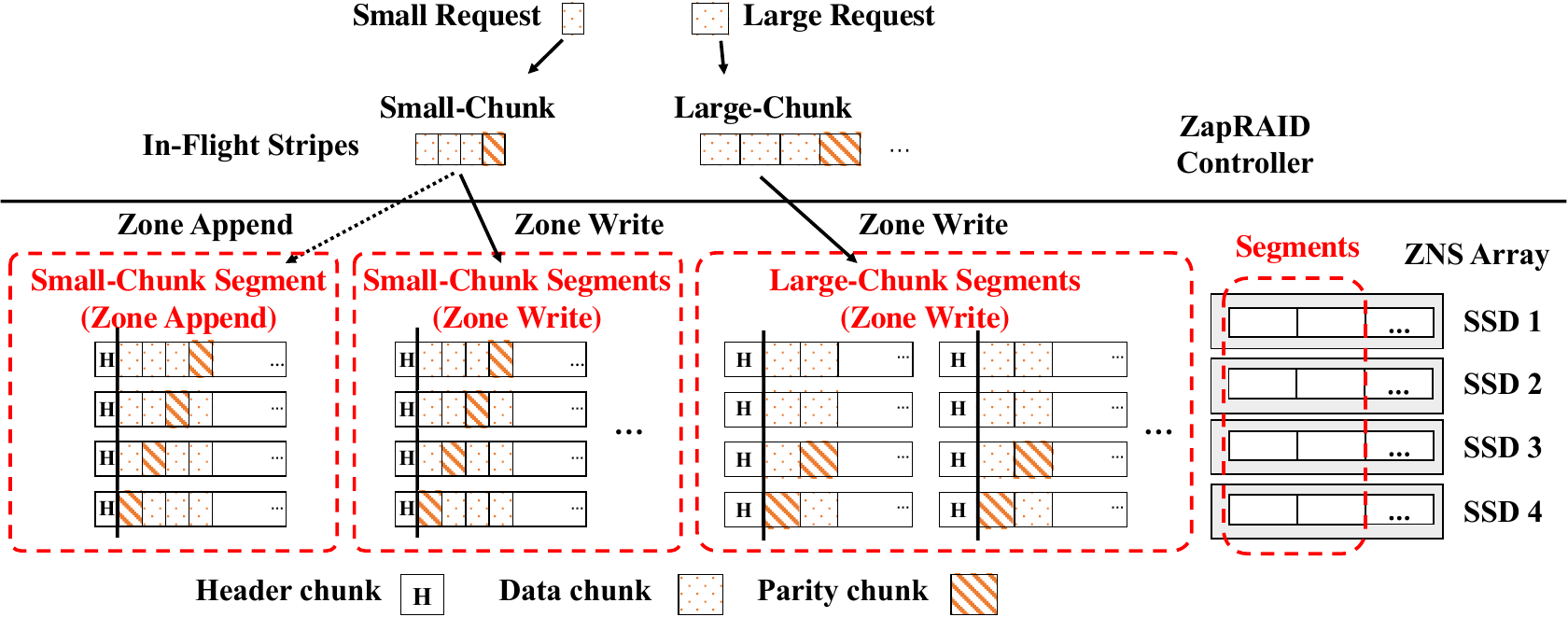}
\vspace{-6pt}
\caption{Hybrid data management of a (3+1)-RAID-5 array with multiple open
segments.}
\label{fig:hybrid}
\vspace{-6pt}
\end{figure}

\para{Segment management.} Figure~\ref{fig:hybrid} shows how \sysname
manages multiple open small-chunk and large-chunk segments.  Let $N_s$ and
$N_l$ be the numbers of open small-chunk segments and open large-chunk
segments, respectively (i.e., the total number of open zones in each ZNS SSD is
$N_s+N_l$). Also, let $C_s$ and $C_l$ be the sizes of a small chunk and a
large chunk, respectively.  All of $N_s$, $N_l$, $C_s$, and $C_l$ are
configurable parameters.  For example, we can set $C_s=8$\,KiB for the small
chunk size and $C_l=16$\,KiB for the large chunk size.  We can also set $N_s >
1$ and $N_l > 1$ to accommodate the real-world workloads that are mixed with
small and large writes (e.g., in cloud block storage \cite{li20}).  

To determine if a write request is issued to a small-chunk or large-chunk
segment, \sysname examines whether the size of the write request is smaller
than a write request size threshold, which we now set as $C_l$.
If so, \sysname assigns the write request to a small-chunk segment; otherwise,
it assigns the write request to a large-chunk segment.  Also, during garbage
collection (\S\ref{sec:impl}), the rewritten data is written to a large-chunk
segment to mitigate I/O overhead. 

\para{Choosing segments from in-flight stripes.} \sysname maintains
in-flight stripes separately for small and large chunks.  It assigns
new written blocks to an in-flight stripe for either small or large
chunks based on the write request size (see above).  When an in-flight stripe
is formed, \sysname forwards the in-flight stripe to an open segment.
Specifically, if the in-flight stripe holds large chunks, \sysname selects an
open large-chunk segment (out of the $N_l$) in a round-robin manner and writes
the in-flight stripe to the selected segment via Zone Write.  On the other
hand, if the in-flight stripe holds small chunks, \sysname further considers
two cases of $N_s$.  If $N_s = 1$, it writes the in-flight stripe to the only
open small-chunk segment via Zone Append.  If $N_s > 1$, it selects an idle
small-chunk segment (out of the $N_s - 1$) in a round-robin manner and writes
the in-flight stripe to the selected segment via Zone Write; however, if there
is no idle small-chunk segment, \sysname selects the only small-chunk segment
reserved for Zone Append and writes the in-flight stripe to the segment via
Zone Append.  

\subsection{Crash Consistency}
\label{subsec:crash}

We show how \sysname maintains consistency after a system crash, in the order
of recovering the consistent states of the segment table, stripes, L2P table,
and compact stripe table (for Zone Append only).  

\para{Recovery of the segment table.} \sysname first identifies the open
and sealed segments in the system, so as to recover the segment table.
Specifically, for each ZNS SSD, \sysname checks the states of all zones.
If a zone is open or full, \sysname retrieves its first block, which belongs
to the header region of a segment (\S\ref{subsec:overview}).  \sysname then
reconstructs the segment table with the information of all possible segments
based on the first blocks of all open and full zones. 

\sysname further checks the validity of each segment.  There are two
possible cases:
\begin{itemize}[leftmargin=*]
\item 
Case~1: the write pointers of all the zones in the segment are equal to or
exceed the header region size. In this case, the segment is a valid segment.
\item
Case~2: the write pointers of some zones are zero. This happens when \sysname
either has created the segment but has only issued writes to some zones,
or has reclaimed the segment and has reset some zones.  In either case,
\sysname simply resets all zones and discards the segment.
\end{itemize}

For each valid segment, if the write pointers of all zones are equal to the
zone capacity, it is a sealed segment; otherwise, it is an open segment.
In particular, if the data region of an open segment reaches its maximum size
but the segment is not yet sealed, \sysname will retrieve all blocks from the
data region to reconstruct the block metadata. It then writes the footer
region and seals the segment. 

\para{Recovery of stripes.} \sysname needs to recover the consistent
states of all stripes by ensuring that each stripe has persisted $k$ data
chunks and $m$ parity chunks.  Specifically, \sysname examines whether each
in-flight stripe has all its $k+m$ chunks persisted. Under the group-based
data layout (note that $G=1$ means the use of Zone Write), \sysname ensures
that only the stripes in the latest stripe group of an open segment can
be the in-flight stripes. It retrieves all the chunks in the latest stripe
group and examines the stripe IDs in their block metadata.  \sysname discards
all partially persisted stripes with fewer than $k+m$ persisted chunks. Recall
that \sysname acknowledges the writes of an in-flight stripe only after
persisting all $k+m$ chunks of the stripe (\S\ref{subsec:overview}), so
discarding the partially persisted stripes does not incur data loss.  Since ZNS
SSDs cannot perform in-place writes to replace the partially persisted chunks,
\sysname rewrites all fully persisted stripes in the segment to a new segment
and reclaims the old segment.

\para{Recovery of the L2P table and the compact stripe table.} After
ensuring the consistency of all stripes, \sysname recovers the L2P table and
the compact stripe table (the latter is for segments that use Zone Append).
Recall that for each sealed segment, \sysname stores the block metadata (i.e.,
the LBA, write timestamp, and stripe ID) of all blocks in the segment in the
footer region, while for each open segment, \sysname also stores the block
metadata in the out-of-band area associated with each block.  To recover the
L2P table and the compact stripe table, \sysname first retrieves the footer
regions of all sealed segments and the data regions of all open segments. It
examines all the block metadata and reconstructs the LBA-to-PBA mappings in
the L2P table and the stripe IDs in the compact stripe table. For the L2P
table, if multiple blocks have the same LBAs, \sysname keeps the one with the
latest write timestamp.

To recover the L2P table when it is offloaded to ZNS SSDs
(\S\ref{subsec:overview}), \sysname extends the above recovery steps for the
L2P table as follows.  During crash recovery, \sysname examines the block
metadata of each block and determines if the block is a user-written block or
a mapping block (based on the least significant bit of the LBA).  If the block
is a user-written block, \sysname reconstructs the LBA-to-PBA mapping in the
L2P table as described above; if the block is a mapping block, since the LBA
in the block metadata refers to the LBA of its first L2P table entry, \sysname
records the LBA-to-PBA mapping of the mapping block in a temporary in-memory
table (if there exist multiple mapping blocks with the same LBA, only the one
with the latest write timestamp is kept). \sysname then examines the
reconstructed L2P table and removes any entry group that has a mapping block
whose latest write timestamp is no smaller than that of any L2P table entry in
the entry group (i.e., the mapping block has the latest L2P entries).
Finally, \sysname reconstructs the in-memory bitmap and the in-memory mapping
table for the on-SSD mapping blocks. 

\subsection{Complete Workflows}
\label{subsec:all}

We explain the complete workflows of writes, normal reads, degraded reads, and
full-drive recovery of \sysname. 

\para{Writes.} To issue a write request (identified by an LBA), \sysname
checks whether the write request is a small or large one based on the write
size, and writes accordingly the data to an in-flight stripe in memory for
small chunks or large chunks (\S\ref{subsec:hybrid}).  If a new data chunk is
available, \sysname issues a Zone Append or Zone Write command to write the
chunk and block metadata based on the chunk's position in the stripe and the
RAID scheme.  When an in-flight stripe contains $k$ data chunks, \sysname
encodes them to generate $m$ parity chunks and their block metadata.  It then
issues a Zone Append or Zone Write command for each of the parity chunks to
its respective zone.  Only after all the chunks of a stripe are persisted,
\sysname updates the L2P table with the corresponding LBAs and PBAs, as well as
acknowledges the completion of the write requests; the in-flight stripe is
also released from memory.  If there are insufficient data chunks to form a
full stripe after a small timeout since the stripe is created (currently set
as 100\,$\mu$s in our prototype, which is close to the median write latency of
16\,KiB chunks (\S\ref{subsec:micro})), \sysname fills the remaining stripe
with zero blocks and invalid LBAs.

\sysname needs to handle two specific cases during writes.  First, if the
written stripe is the first stripe in a stripe group, \sysname should ensure
that all the stripes in the previous stripe group are completely persisted;
if the stripe group is also the first one of a segment, the header region of
the segment should be completely persisted.  Second, if the stripe group is
also the last one in the data region, \sysname writes the block metadata of
all blocks in the segment into the footer region in the background.
It also creates a new open segment and writes the segment information to the
header region (without waiting for the completion of the writing of
the footer region in the previous segment), so that the new open segment can
serve new writes. 

\para{Reads.} To read a block (identified by an LBA), \sysname queries the
L2P table for its PBA (i.e., the segment ID, the drive ID, and the offset in
the respective zone). Given the segment ID and drive ID, \sysname locates the
zone from the segment table and retrieves the block from the specified offset
of the zone.

\para{Degraded reads.} Suppose that \sysname performs a degraded read to
a lost block.  Based on the PBA of the requested block, \sysname checks
whether the segment applies Zone Append or Zone Write.  For Zone Append,
\sysname queries the compact stripe table to find out the stripe ID of the
requested block. Since all available chunks of the same stripe must reside in
the same stripe group under the group-based data layout, \sysname searches for
the offsets of the $k$ available chunks in the stripe group from the compact
stripe table; note that the search is efficient due to the limited group size
(\S\ref{subsec:micro}).  It then reads the available chunks from the other
zones, decodes the lost chunk, and retrieves the requested block.  For Zone
Write, \sysname reads the available chunks from other zones at the same offset
to recover the chunk and hence the requested block. 

\para{Full-drive recovery.} When a drive fails, \sysname recovers the
lost data into a new drive.  It first identifies the segments that contain the
lost zones in the failed drive by examining the segment-to-zones mappings in
the header regions of all stored segments. For each lost zone, \sysname
retrieves all available zones in the same segment from other available drives
into memory, and reconstructs the stripe groups that cover the lost zone.  It
examines the block metadata of the available chunks and identifies the
chunks from the same stripe. It then decodes the lost chunks for each stripe
independently.  After repairing all the stripes in a segment, \sysname writes
the recovered zone to the new drive. 

\subsection{Discussion}

We now discuss some open issues of our \sysname design.

\para{Choice of the write request size threshold in hybrid data
management.} Recall that \sysname compares the size of a write request against
a threshold to determine if the write request is assigned to a small-chunk or
large-chunk open segment (\S\ref{subsec:hybrid}).  Currently, the threshold is
set to $C_l$.  The choice of the threshold poses different performance
trade-offs.  A larger threshold implies that more writes will go to
small-chunk segments.  Full small-chunk stripes are formed faster, and hence
small writes are also acknowledged faster (i.e., lower response time).
However, the write throughput is lower since writing small-chunk segments is
slower than writing large-chunk segments (based on the results in Figure~2 for
different write request sizes).  For simplicity, we assume that the threshold
is equal to $C_l$ (i.e., if a write request size is smaller than
$C_l$, it is assigned to a small-chunk segment; otherwise, it is assigned to a
large-chunk segment).

\para{Relationships between the \sysname design and the ZNS SSD model.}
While our motivating experiments and evaluation are based on the ZN540 model
\cite{zn540}, our \sysname design does not heavily depend on a specific ZNS
SSD model.  For example, some of our key design features, such as the
group-based data layout and crash recovery management, apply to all ZNS SSD
models. 

\sysname is mainly based on two design principles: (i) Zone Append outperforms
Zone Write in one or a few open zones for small writes, and (ii) a combination
of Zone Append and Zone Write can be used for hybrid data management when there
are multiple open zones.  We note that \sysname now only uses Zone Append for
a single open segment in hybrid data management since the improvement of Zone
Append is limited for a large number of open zones, and we conjecture that the
reason is due to the computational overhead of the firmware.  Nevertheless, if
the computational overhead can be mitigated in future ZNS SSD models, \sysname
can issue Zone Append to more open segments to better exploit intra-zone
parallelism; this can be readily achieved with parameter reconfigurations
based on our current design. 

Our \sysname design is also applicable to different zone capacities of ZNS
SSDs. For smaller zone capacities, the data and footer regions become smaller,
yet the group-based data layout remains applicable. For example, when the zone
capacity is 96\,MiB \cite{min23} (or equivalently, 24,576 blocks), for 4\,KiB
chunks, the header, data, and footer regions occupy 1~block, 24,455~blocks,
and 120~blocks, respectively (based on the calculation of
\S\ref{subsec:overview}). If the stripe group size is 256, there are still
many stripe groups (i.e., 96~groups) for intra-zone parallelism.  

Note that the ZN540 model has been widely studied by a large body of prior
work in the literature (e.g., \cite{bergman22, kim23, lee23, seo23,
doekemeijer23}) and is a representative platform for current ZNS SSD research. 

\section{Implementation}
\label{sec:impl}

Since ZNS SSDs offer low latency and high throughput, the overhead of
the Linux kernel I/O stack (e.g., high latencies of interrupts, locks, and
system calls) is no longer negligible \cite{lee19}. To reduce I/O latency, we
implement \sysname using the Storage Performance Development Kit (SPDK)
\cite{spdkbdev}.  SPDK supports polling-based and lockless I/Os. It also
avoids system calls by moving I/O drivers to the user space.

Specifically, we prototype \sysname as a user-space block device module in
C++ with around 9.7\,K\,LoC based on SPDK.  \sysname exports a block
interface to be compatible with general applications. Our prototype
decomposes request handling into seven handlers, namely {\em dispatch}, {\em
device I/O}, {\em completion}, {\em indexing}, {\em encoding}, {\em segment
state tracking}, and {\em cleaning}.

\para{Dispatch handler.} The dispatch handler is responsible for
processing write requests and updating the header and footer regions.  For
write requests, the dispatch handler assigns the written blocks to segments
and submits the I/O events for the written blocks to the device I/O handler.
If a block fills a stripe in a segment, the dispatch handler also sends a
parity generation event to the encoding handler, which computes parity chunks
and submits the I/O events for the parity chunks to the device I/O handler.
Recall that the writes of the first stripe of a stripe group need to wait for
the previous stripe group to be completely persisted (\S\ref{subsec:all}); in
this case, the dispatch handler will suspend the request and retry later until
the segment state is updated by the segment state tracking handler (see below). 

For read requests, the dispatch handler first sends an L2P table query event to
the indexing handler, which asynchronously notifies the dispatch handler to issue 
the read requests upon retrieving PBAs for the requested blocks. The dispatch
handler then submits the I/O events to the device I/O handler to retrieve the
requested blocks from storage.

\para{Device I/O handler.} The device I/O handler is
responsible for submitting block I/O requests to the drives. It receives I/O
events from other handlers and submits the corresponding I/O requests
to the drives. It is also responsible for checking the completion of each block
I/O request from the drives and notifying the completion handler (see
below).

\para{Completion handler.} The completion handler tracks the completion state
of each I/O request.  It receives the notification of completion of a block
I/O event from the device I/O handler.  If all the I/O events for a request are
completed, the completion handler notifies the upper-layer application.  The
application can register a callback function for each request.  For write
requests, the completion handler first sends an L2P table update event to the
indexing handler, which triggers the callback function after updating the L2P
table.  For read requests, the completion handler directly triggers the callback
function. 

If a block is lost or unavailable during a read request, the completion handler
issues a degraded read to the block.  It examines the compact stripe table to
identify the PBAs of the available blocks of the same stripe. It then sends
the I/O events to the device I/O handler to retrieve the available blocks.  After
all available blocks are retrieved, the completion handler decodes the lost block
and marks the read request as completed.

\para{Indexing handler.} The indexing handler is responsible for the L2P table
maintenance. It receives L2P table query events from the dispatch handler and
returns the PBAs of requested blocks for read requests.  It also receives L2P
table update events from the completion handler and the cleaning handler.  
If an L2P table update event is from the completion handler, it acknowledges the
write requests to the application on behalf of the completion handler after the
L2P table is updated.  In addition, the indexing handler is responsible for the
offloading of L2P table entries if the L2P table size reaches the memory
size limit (\S\ref{subsec:overview}). 

\para{Encoding handler.} The encoding handler is responsible for parity
generation.  When it receives a notification from the dispatch handler that a
stripe has $k$ data chunks, it generates parity chunks. It then submits I/O
events for the parity chunks to the device I/O handler.  We handle parity chunk
generation outside of the normal I/O requests to mitigate interference on
normal I/O requests.

\para{Segment state tracking handler.} The segment state tracking handler
examines whether the system needs to write the header region or the footer
region of a segment, or whether it has persisted all stripes in a
stripe group (\S\ref{subsec:all}). It periodically (every 1\,$\mu$s in our
prototype) examines the segment states.  When the segment state needs to be
updated, it also sends the corresponding I/O events to the device I/O
handler. 

\para{Cleaning handler.} The cleaning handler is responsible for
reclaiming the space of stale blocks. It periodically (every 1\,$\mu$s in our
prototype) examines the available space and triggers garbage collection if the
available space drops below a fixed threshold.  It selects one sealed
segment that has the most stale blocks and rewrites its valid blocks by
submitting block I/O events to the device I/O handler.  Currently, our
prototype tracks the validity of all blocks (including both user-written
blocks and mapping blocks) for each segment via an in-memory bitmap (whose
content can also be derived from both the L2P table for user-written blocks
and the LBA-to-PBA table for mapping blocks) and collects the LBAs of valid
blocks from the block metadata.  When all rewrites are completed, the dispatch
handler sends the LBAs of all rewritten blocks, along with their old and new
PBAs, to the indexing handler for updating the L2P table.

In this work, we do not further consider sophisticated garbage
collection algorithms, which need to address the key design issues of (i) when
to trigger garbage collection, (ii) which segment to clean for garbage
collection, and (iii) where to rewrite the valid data.  We argue that such
design issues are orthogonal to our current focus on leveraging Zone Append
and Zone Write for a high-performance RAID design.  We pose the design of new
garbage collection algorithms as future work.

\para{Thread assignment.} We assign the seven handlers to SPDK threads,
which exchange events via \texttt{spdk\_thread\_send\_msg} and event
polling.  Currently, we co-locate the dispatch handler, completion handler,
segment state tracking handler, and cleaning handler in the main SPDK thread.
We also assign the device I/O handler, indexing handler, and encoding handler
separately to three different SPDK threads for parallelism, such that our
prototype can saturate the write bandwidth of our testbed. 

\section{Evaluation}
\label{sec:eval}

We evaluate \sysname through real-application experiments,
trace-driven experiments, and microbenchmarks.

\subsection{Methodology}
\label{subsec:settings}

We use a server that runs Ubuntu 22.04 LTS with Linux kernel 5.15. It has a
16-core Intel Xeon Silver 4215 2.5\,GHz CPU and 96\,GiB DRAM. It is attached
with four 4\,TiB Western Digital Ultrastar DC ZN540 ZNS SSDs \cite{zn540}.
Each SSD has 3,690 zones, with a zone capacity of 1,077\,MiB each.  We format
each SSD with a logical block size of 4\,KiB and a block metadata size (in the
out-of-band area) of 64~bytes.  By default, we focus on (3+1)-RAID-5, yet we
also consider other RAID schemes and a larger-scale emulated ZNS SSD array
based on FEMU \cite{li18} (\S\ref{subsec:micro}). 

We mainly consider two baselines: {\em ZoneWrite-Only} and {\em
ZoneAppend-Only}. ZoneWrite-Only always uses Zone Write for all writes (i.e.,
$G=1$), while ZoneAppend-Only always uses Zone Append for all writes and does
not employ a group-based data layout (i.e., $G=S$) (\S\ref{subsec:group}).
For fair comparisons, when there are multiple open segments,
ZoneWrite-Only and ZoneAppend-Only also write small chunks to small-chunk
segments and large chunks to large-chunk segments, only using Zone Write and
Zone Append, respectively.

We also compare \sysname with a simplified version of RAIZN \cite{kim23},
which we implement as a user-space block device based on SPDK (referred
to as {\em RAIZN-SPDK}).  Note that the open-source
version of RAIZN \cite{raizn} is currently implemented as a kernel device
mapper and exports a ZNS interface as a logical ZNS SSD volume, so
we re-implement RAIZN as a user-space block device for fair comparisons with
\sysname. Specifically, the original RAIZN reserves at least three additional
metadata zones (in addition to regular zones for data storage) in each drive,
including one zone for appending partial parity updates before acknowledging
each write request, one zone for appending metadata to ensure write atomicity,
and at least one zone for the garbage collection of metadata zones.  It caches
the written data in an in-memory stripe buffer and writes the partial
parity updates to the partial-parity-dedicated metadata zone, so as to support
crash recovery even if a system crash happens before the whole stripe is
written.  When the stripe buffer is full, RAIZN generates and writes the
complete parity chunks to the regular zones, and discards the corresponding
partial parity updates.  RAIZN differs from ZoneWrite-Only, ZoneAppend-Only,
and \sysname in that it acknowledges each write request after the
corresponding partial parity updates are appended to the dedicated metadata
zone, while the other three schemes leverage concurrent write requests to form
full stripes before acknowledging the write requests.

In our RAIZN-SPDK implementation, we focus on achieving high performance. We
simplify our implementation by removing the metadata zones reserved for write
atomicity and garbage collection.  RAIZN-SPDK also reserves two metadata zones
for high performance.  Specifically, it writes partial parity updates to one
of the two metadata zones.  When the metadata zone becomes full, RAIZN-SPDK
resets the zone and	writes the partial parity updates to another metadata zone
in parallel.  For high performance, RAIZN-SPDK uses Zone Append to write partial
parity updates to a metadata zone as described in \cite{kim23}.  When there
are multiple open segments, RAIZN-SPDK uses Zone Write to write small chunks
to small-chunk segments and large chunks to large-chunk segments, as in
ZoneWrite-Only, while using Zone Append to write partial parity updates of all
segments to a metadata zone.

We also attempt to fairly compare \sysname with the conventional RAID
array that runs on traditional block-interface SSDs.  However, the firmware
on our ZN540 SSDs is locked and cannot be directly modified, so we cannot
make comparisons on the same hardware as in \cite{kim23}.  Thus, we leverage
FEMU-emulated SSDs \cite{li18} for comparisons and build the conventional RAID
array with {\tt mdadm} \cite{mdadm} (see Exp\#10 for details).

Our experiments consider the configurations with a single open segment
and multiple open segments.  If there is a single open segment, \sysname
always issues writes via Zone Append and applies the group-based data layout;
if there are multiple open segments, \sysname enables hybrid segment
management and issues writes via a combination of Zone Append and Zone Write
(the exact configurations are further explained in individual experiments).

For each experiment, we report the average results over five runs and include
the 95\% confidence intervals based on Student's t-distribution (note that
some confidence intervals may be invisible due to negligible deviations across
runs). 

Our evaluation is driven by the following research questions:
\begin{itemize}[leftmargin=*,topsep=0pt]
\item 
How does \sysname perform on real applications in terms of writes, degraded
reads, and memory usage for index management, when compared to ZoneWrite-Only
and ZoneAppend-Only?  (RQ1)
\item 
How does \sysname perform under realistic cloud storage traces that are
composed of a mix of small and large writes? (RQ2)
\item 
How does \sysname perform across different aspects under microbenchmarks,
including writes, normal reads, degraded reads, crash recovery, and full-drive
recovery? (RQ3)
\end{itemize}

\subsection{Real-Application Experiments (RQ1)}
\label{subsec:app}

We start our evaluation with real applications to demonstrate the
end-to-end performance benefits of \sysname, in comparison with ZoneWrite-Only
and ZoneAppend-Only (similar benchmarks are also reported in \cite{kim23}). We
exclude RAIZN-SPDK, as it shows poor performance from our microbenchmarks
(\S\ref{subsec:micro}).

To host Linux applications, we export our \sysname prototype as an
NVMe-over-Fabrics (NVMe-oF) target with TCP transport \cite{spdknvmeof} and
deploy the NVMe-oF host in the same testbed machine, so that the prototype
serves as a Linux block device.  We allocate the block device with a logical
space capacity of 1\,TiB, format the block device as an EXT4 file system, and
run different types of applications atop the file system. 
Since the application write traffic may exceed the logical capacity, we 
reserve an additional 25\% of the physical capacity for garbage collection.  

We evaluate \sysname's performance on both single-segment and multi-segment
settings. For multi-segment settings, we vary the numbers of small-chunk and
large-chunk segments (\S\ref{subsec:hybrid}).  We fix the small- and
large-chunk sizes at $C_s=8$\,KiB and $C_l=16$\,KiB, respectively, and keep
the total number of open segments as four (i.e., $N_s+N_l=4$).  Here, we focus
on four representative settings of open segments: (i) a single open segment
with a chunk size of 4\,KiB, (ii) a single open segment with a chunk size of
16\,KiB, (iii) four open segments with $(N_s,N_l)=(2,2)$ (i.e., two
small-chunk segments and two large-chunk segments), and (iv) four open
segments with $(N_s,N_l)=(1,3)$ (i.e., one small-chunk segment and three
large-chunk segments).  We also present extensive evaluation results of
various open-segment configurations in microbenchmarks in
\S\ref{subsec:micro}.

\para{Workloads.} 
We evaluate three types of application workloads: file writes, key-value
workloads, and concurrent workloads.  

\emph{File writes.} 
We evaluate \sysname in file writes using Sysbench (v1.0.20) \cite{sysbench}
to generate continuous write patterns. We first create 128 files of 800\,MiB
each (totaling 100\,GiB). We then use Sysbench to issue continuous write
requests with request sizes of 4\,KiB, 8\,KiB, 16\,KiB, 32\,KiB, and 64\,KiB. 
For each request size, we run the workload for 10~minutes.  Unless otherwise
specified, we use one thread to issue writes in asynchronous mode with direct
I/O, and each run ends with an \texttt{fsync} to ensure data persistence.
Before each run, we clear and reformat the block device as an EXT4 file
system.  This workload stresses small, frequent writes and highlights both the
throughput and metadata efficiency of different schemes.  

\emph{Key-value store workloads.} We evaluate \sysname when it hosts RocksDB
(v9.7.2) \cite{rocksdb} and use its benchmarking tool
\texttt{db\_bench} \cite{dbbench} to issue key-value operations. We focus on
two settings of the \texttt{fillrandom} workload: (i) the default
\texttt{fillrandom}, which inserts randomly ordered keys into an empty
database and triggers background compaction, and (ii) \texttt{fillrandom} with
compaction disabled (bulk loading), which sequentially writes data without
compaction overhead. We configure the database with approximately 1.64~million
key-value pairs, each with a value size of 64\,KiB, and disable compression.
We use eight threads per client and run a total of eight clients concurrently,
resulting in 64 writer threads in total. The workload writes about 100\,GiB of
data in total, evenly distributed across all clients. For each run, we reset
the volume and then issue one of the two \texttt{fillrandom} workloads. These
two workload settings capture both the compaction overhead of LSM-tree
maintenance and the pure bulk-loading behavior without compaction.  

\emph{Concurrent workloads.} We combine two types of write-intensive workloads
with different write patterns to evaluate \sysname. Specifically, we use
Sysbench to generate small file writes, while simultaneously hosting MySQL and
running the \texttt{oltp\_write\_only} benchmark, which issues large writes
with fewer than 10\% reads. This setting emulates a virtualized environment in
which multiple applications coexist in an SSD RAID array \cite{colgrove15}.
For the small file writes, we use Sysbench to continuously issue 4\,KiB write
requests. For the \texttt{oltp\_write\_only} workload, we use Sysbench to
populate the MySQL database with 8 tables and 10~million rows, and then
execute \texttt{DELETE}, \texttt{INSERT}, and \texttt{UPDATE} queries using 64
threads. Before each run, we clear and reformat the block device as an EXT4
file system, and then execute both benchmarks simultaneously for 10~minutes.  

\begin{figure*}[!t]
\centering
\setlength{\tabcolsep}{3pt}
\begin{tabular}{@{}cccc@{}}
\multicolumn{4}{c}{
\includegraphics[width=0.53\textwidth]{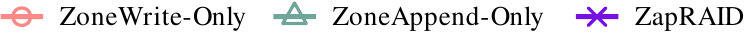}
}\\
\includegraphics[width=0.235\textwidth]{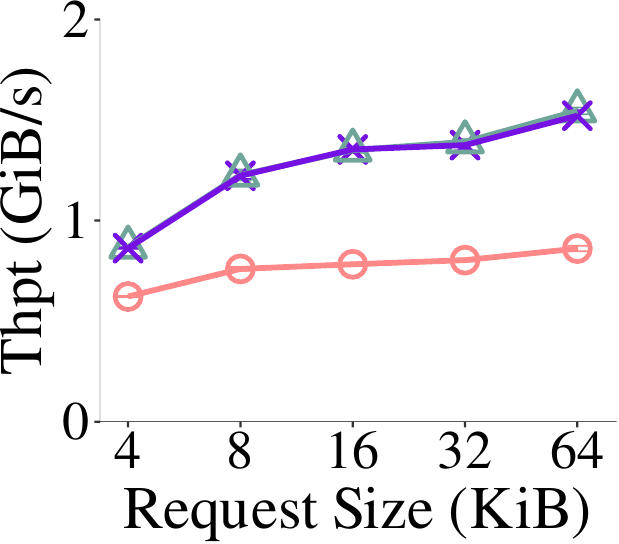} &
\includegraphics[width=0.235\textwidth]{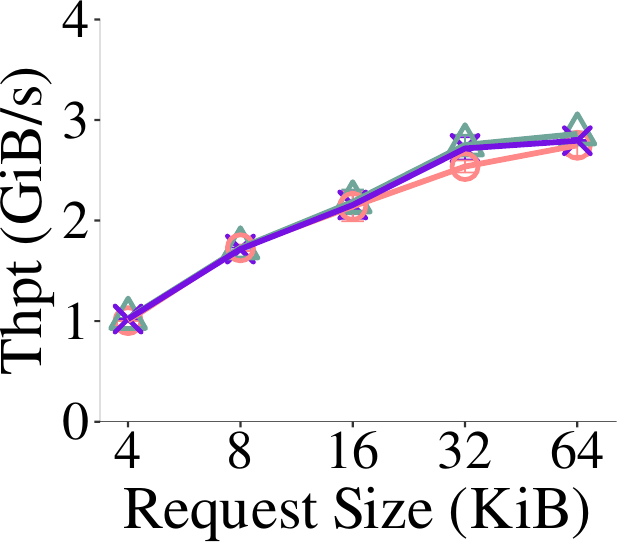} &
\includegraphics[width=0.235\textwidth]{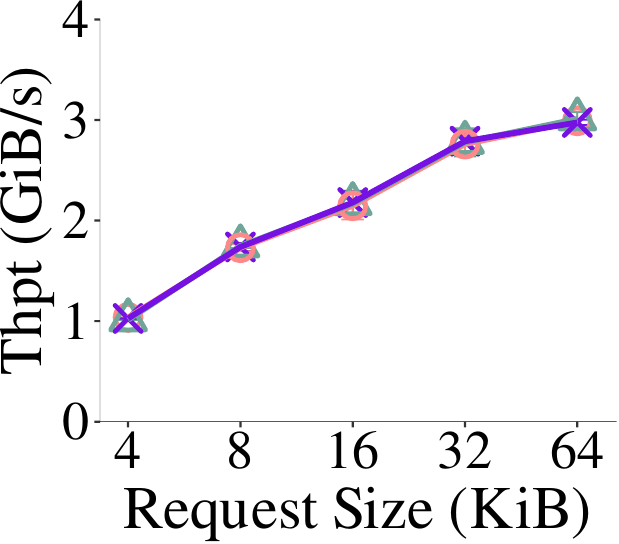} &
\includegraphics[width=0.235\textwidth]{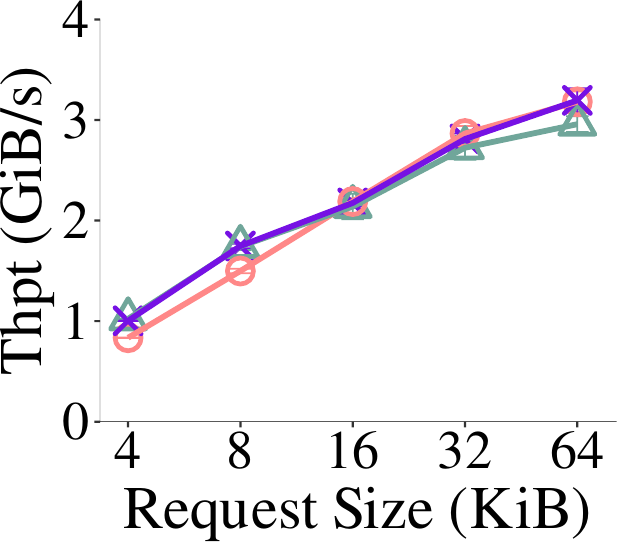} \\
\parbox[t]{0.235\textwidth}{\centering \small (a) File writes, single (4\,KiB)} &
\parbox[t]{0.235\textwidth}{\centering \small (b) File writes, single (16\,KiB)} &
\parbox[t]{0.235\textwidth}{\centering \small (c) File writes, multiple $(N_s,N_l)=(2,2)$} &
\parbox[t]{0.235\textwidth}{\centering \small (d) File writes, multiple $(N_s,N_l)=(1,3)$} 
\vspace{6pt}\\
\multicolumn{4}{c}{
\includegraphics[width=0.60\textwidth]{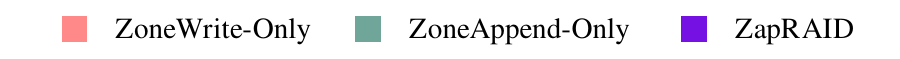}
}\\
\includegraphics[width=0.235\textwidth]{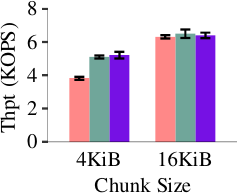} &
\includegraphics[width=0.235\textwidth]{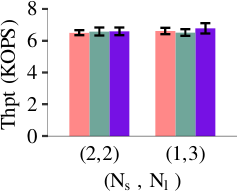} &
\includegraphics[width=0.235\textwidth]{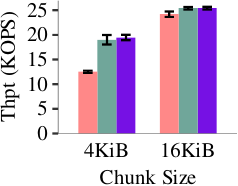} &
\includegraphics[width=0.235\textwidth]{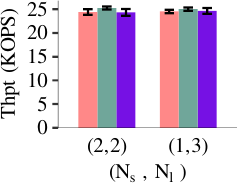} \\
\parbox[t]{0.235\textwidth}{\centering \small (e) Key-value (with compaction), single} &
\parbox[t]{0.235\textwidth}{\centering \small (f) Key-value (with compaction), multiple} &
\parbox[t]{0.235\textwidth}{\centering \small (g) Key-value (bulk load, no compaction), single} &
\parbox[t]{0.235\textwidth}{\centering \small (h) Key-value (bulk load, no compaction), multiple}
\vspace{6pt}\\
\includegraphics[width=0.235\textwidth]{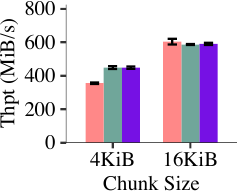} &
\includegraphics[width=0.235\textwidth]{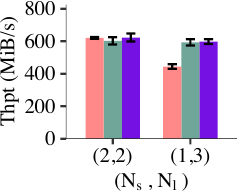} &
\includegraphics[width=0.235\textwidth]{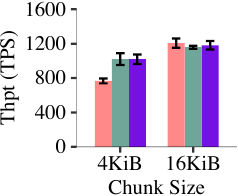} &
\includegraphics[width=0.235\textwidth]{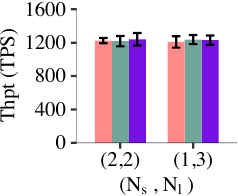} \\
\parbox[t]{0.235\textwidth}{\centering \small (i) Concurrent: file writes
(4\,KiB), single} & \parbox[t]{0.235\textwidth}{\centering \small (j)
Concurrent: file writes (4\,KiB), multiple} &
\parbox[t]{0.235\textwidth}{\centering \small (k) Concurrent:
\texttt{oltp\_ write\_only}, single} &
\parbox[t]{0.235\textwidth}{\centering \small (l) Concurrent:
\texttt{oltp\_ write\_only}, multiple}
\end{tabular}
\vspace{-6pt}
\caption{Exp\#1 (Write performance): file writes (figures~(a)-(d)), key-value
store workloads (figures~(e)-(h)), and concurrent workloads (figures~(i)-(l)).}
\label{fig:realapp_all}
\end{figure*}

\para{Exp\#1 (Write performance).} We first evaluate the write
performance of \sysname and compare it against ZoneWrite-Only and
ZoneAppend-Only under different open-segment settings.

\emph{File writes.} Figures~\ref{fig:realapp_all}(a)-\ref{fig:realapp_all}(d)
present the throughput results of sequential file writes under different
open-segment settings. For a single open segment with 4\,KiB chunks
(Figure~\ref{fig:realapp_all}(a)), \sysname achieves throughput comparable to
ZoneAppend, while outperforming ZoneWrite-Only by 38.6\%. At larger request
sizes, the performance gap between \sysname and ZoneWrite-Only increases
significantly, reaching 60.9\%, 73.0\%, 71.2\%, and 76.8\% at 8\,KiB, 16\,KiB,
32\,KiB, and 64\,KiB, respectively.  The gains mainly stem from \sysname's
ability to exploit intra-zone parallelism under a small chunk size, which is
consistent with our observation in Figure~\ref{fig:motivation}(a) in
\S\ref{subsec:vs} for a single open segment with 4\,KiB chunks.  

At small request sizes (e.g., 4\,KiB), \sysname's improvement is relatively
modest due to the system overhead from file system metadata updates, as well as
kernel I/O stack and NVMe-oF's TCP stack operations. As the request size
becomes larger, such overhead becomes amortized.  Since the chunk size is
4\,KiB, large requests are still issued to small chunks. \sysname can deliver
progressively larger throughput gains over ZoneWrite-Only via intra-zone
parallelism.  

\begin{table}[t]
\renewcommand{\arraystretch}{1.15}
\footnotesize
\centering
\begin{tabular}{|c|c|l|c|}
\hline
\textbf{Settings} & \textbf{Request sizes} & \makecell[c]{\textbf{Findings}} &
\textbf{Reasons} \\ \hline
\multirow{2}{*}{\makecell[c]{Single segment,\\ 4\,KiB chunks}}
  & 4-8\,KiB & Modest improvements over ZoneWrite-Only &
  Reasons A and B \\ \cline{2-4}
  & 16-64\,KiB & Significant improvements over ZoneWrite-Only &  
  Reason A \\ \hline
\multirow{2}{*}{\makecell[c]{Single segment,\\ 16\,KiB chunks}}
  & 4-8\,KiB & \multirow{2}{*}{Marginal difference} & \multirow{2}{*}{Reason C} \\ \cline{2-2}
  & 16-64\,KiB &  &  \\ \hline
\multirow{2}{*}{\makecell[c]{Multiple segments,\\ $(N_s,N_l)=(2,2)$}}
  & 4-8\,KiB & Marginal difference & Reason D \\ \cline{2-4}
  & 16-64\,KiB & Marginal difference & Reason C \\ \hline
\multirow{2}{*}{\makecell[c]{Multiple segments,\\ $(N_s,N_l)=(1,3)$}}
  & 4-8\,KiB & Modest improvements over ZoneWrite-Only & Reasons A and B \\
	  \cline{2-4}
  & 16-64\,KiB & Marginal difference & Reason C \\ \hline
\multicolumn{4}{|l|}{%
\textbf{Reason~A:} Exploiting intra-zone parallelism with small chunks; see
	Figures~\ref{fig:motivation}(a) and \ref{fig:motivation}(b) in
	\S\ref{subsec:vs}.}\\
\multicolumn{4}{|l|}{%
\textbf{Reason~B:} System overhead of file system metadata updates, I/O
	stacks, and NVMe-oF's TCP stacks.}\\
\multicolumn{4}{|l|}{%
\textbf{Reason~C:} Large writes have similar throughput in Zone Append and
	Zone Write; see Figure~\ref{fig:motivation}(c) in \S\ref{subsec:vs}.}\\ 
\multicolumn{4}{|p{5in}|}{%
\textbf{Reason~D:} Small writes are distributed across zones and have similar
	throughput in Zone Append and Zone Write; see
	Figures~\ref{fig:motivation}(a) and \ref{fig:motivation}(b) in
	\S\ref{subsec:vs}.}\\ 
\hline
\end{tabular}
\vspace{3pt}
\caption{Exp\#1 (Write performance). Explanations of the throughput results 
for file writes across different request sizes and segment settings.}
\label{tab:exp1_filewrite}
\end{table}

For a single open segment with 16\,KiB chunks and multiple open segments with
$(N_s,N_l)=(2,2)$ (Figures~\ref{fig:realapp_all}(b) and
\ref{fig:realapp_all}(c), respectively), the performance differences among the
three schemes are small.  For the former, the 4\,KiB writes are absorbed into
16\,KiB chunks and issued as large writes, which are handled comparably by
Zone Append and Zone Write (Figure~\ref{fig:motivation}(c) in
\S\ref{subsec:vs}); for the latter, the 4\,KiB writes are distributed across
two zones, where Zone Append and Zone Write deliver similar performance
(Figure~\ref{fig:motivation}(a) in \S\ref{subsec:vs}).  For multiple open
segments with $(N_s,N_l)=(1,3)$ (Figure~\ref{fig:realapp_all}(d)), \sysname
shows a modest advantage at small request sizes (subject to the system
overhead), outperforming ZoneWrite-Only by 20.2\% and 16.6\% at 4\,KiB and
8\,KiB, respectively, while the difference narrows at large request sizes
since large writes yield comparable throughput across schemes. 

Table~\ref{tab:exp1_filewrite} summarizes the reasons above to provide a
consolidated view across different request sizes and open segment settings.

Note that across all settings
(Figures~\ref{fig:realapp_all}(a)-\ref{fig:realapp_all}(d)), small requests
of 4\,KiB and 8\,KiB can only achieve the throughput of at most
1,061.50\,MiB/s and 1,789.36\,MiB/s, respectively.  The throughput is
constrained by the overhead of file system metadata updates, kernel I/O
stack operations, and TCP stack operations in NVMe-oF. In Exp\#1, each request
must traverse both the file system and TCP/IP stacks, which incur substantial
processing overhead. In contrast, Exp\#5 (\S\ref{subsec:micro}) issues
requests directly to the SPDK target without traversing the TCP stack, and
the throughput can reach up to 1,599\,MiB/s and 3,012\,MiB/s under 4\,KiB and
8\,KiB writes, respectively. The TCP overhead in NVMe-oF, even with TCP
loopback, has also been reported by the SPDK community \cite{spdk-ublk}.  Such
overhead becomes less significant for large request sizes.

\emph{Key-value store workloads.}
Figures~\ref{fig:realapp_all}(e)-\ref{fig:realapp_all}(h) show the throughput
results of RocksDB using \texttt{db\_bench} under two configurations: the
\texttt{fillrandom} workload with compaction enabled and the bulk-loading
workload without compaction.  

Under \texttt{fillrandom} with compaction (Figures~\ref{fig:realapp_all}(e)
and \ref{fig:realapp_all}(f)), \sysname outperforms ZoneWrite-Only by 41.7\%
under a single open segment with 4\,KiB chunks, while the three schemes perform 
nearly the same as under a single open segment with 16\,KiB chunks
(Figure~\ref{fig:realapp_all}(e)).  For multiple open segments with
$(N_s,N_l)=(2,2)$ and $(N_s,N_l)=(1,3)$, the throughput difference across the
three schemes remains small (Figure~\ref{fig:realapp_all}(f)).

For bulk loading without compaction (Figures~\ref{fig:realapp_all}(g) and
\ref{fig:realapp_all}(h)), \sysname shows larger benefits at small request
sizes. It achieves 51.7\% higher throughput than ZoneWrite-Only under a single
open segment with 4\,KiB chunks, while the advantage narrows to 5.7\% under a
single open segment with 16\,KiB chunks. As with the compaction-enabled 
workload, the multiple open segments with $(N_s,N_l)=(2,2)$ and
$(N_s,N_l)=(1,3)$ yield comparable throughput across all schemes. Since
compaction is disabled during bulk loading, read overheads are eliminated,
making the performance improvement of \sysname more pronounced. 

We observe that the performance gaps across the three schemes are small for the
configurations with a large chunk size (e.g., 16\,KiB) or sufficient open
segments (e.g., four).  RocksDB exhibits both small writes (e.g., write-ahead
log (WAL) writes) and large writes (e.g., flushes and compactions), while
large writes dominate.  Both Zone Write and Zone Append primitives are
expected to have close performance in such configurations (see
Figure~\ref{fig:motivation} in \S\ref{subsec:vs}).

\emph{Concurrent workloads.} Figures~\ref{fig:realapp_all}(i) and
\ref{fig:realapp_all}(j) show the write throughput results for small file
writes in concurrent workloads. \sysname achieves 26.3\% and 34.5\% higher
throughput than ZoneWrite-Only under a single open segment with 4\,KiB chunks
(Figure~\ref{fig:realapp_all}(i)) and four open segments with
$(N_s,N_l)=(1,3)$ (Figure~\ref{fig:realapp_all}(j)), respectively.  In both
settings, all small writes are directed to the single small-chunk segment.
\sysname outperforms ZoneWrite-Only by issuing Zone Append.  On the other
hand, for a single open segment with 16\,KiB chunks and four open segments
with $(N_s,N_l)=(2,2)$, all three schemes have similar performance, due to the
same reasons as explained in the file writes workloads (see above). 

Figures~\ref{fig:realapp_all}(k) and \ref{fig:realapp_all}(l) depict the
throughput of the \texttt{oltp\_write\_only} benchmark in transactions per
second (TPS). For a single open segment with 4\,KiB chunks
(Figure~\ref{fig:realapp_all}(k)), \sysname achieves 32.9\% higher throughput
than ZoneWrite-Only and has similar throughput to ZoneAppend-Only. For other
open-segment settings, all three schemes show similar performance.  In
{\tt oltp\_write\_only}, where large writes dominate, we expect
similar performance across all three schemes for the configurations with a
large chunk size and sufficient open segments (see the explanations above for
key-value store workloads). 

\para{Exp\#2 (Degraded reads).} We evaluate degraded read performance by
first loading data into the system, failing one disk to emulate degraded mode,
and finally issuing read requests based on the file writes and key-value store
workloads.  We consider two workloads, derived from the file writes and
key-value store workloads, respectively: (i) Sysbench with \texttt{readrandom}
and (ii) \texttt{db\_bench} with \texttt{readrandom}.  For the former, we load
a 100\,GiB dataset using Sysbench, fail one device, and perform degraded
random reads continuously for 10~minutes using Sysbench in \texttt{readrandom}
mode.  For the latter, we use \texttt{db\_bench} to first load a 100\,GiB
database with 64\,KiB values, fail one device, and then issue 20,000 degraded
\texttt{readrandom} requests using \texttt{db\_bench}. 

\begin{figure}[!t]
\centering
\setlength{\tabcolsep}{3pt}
\begin{tabular}{@{}cccc@{}}
\multicolumn{4}{c}{
\includegraphics[width=0.60\textwidth]{figs/exp_realapp/write_dbbench_OPS_withcompaction/pdf/ops_legend.pdf}
}\\
\includegraphics[width=0.23\textwidth]{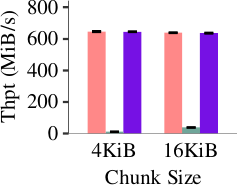} &
\includegraphics[width=0.23\textwidth]{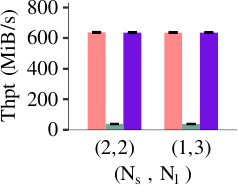} &
\includegraphics[width=0.23\textwidth]{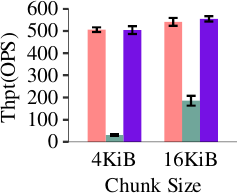} &
\includegraphics[width=0.23\textwidth]{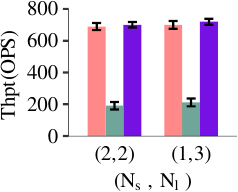} \\
\parbox[t]{0.23\textwidth}{\centering \small (a) Sysbench, single} &
\parbox[t]{0.23\textwidth}{\centering \small (b) Sysbench, multiple} &
\parbox[t]{0.23\textwidth}{\centering \small (c) \texttt{db\_bench}, single} &
\parbox[t]{0.23\textwidth}{\centering \small (d) \texttt{db\_bench}, multiple}
\end{tabular}
\vspace{-6pt}
\caption{Exp\#2 (Degraded reads): Sysbench (figures~(a) and (b)) and
\texttt{db\_bench} (figures~(c) and (d)).}
\label{fig:degraded_reads}
\end{figure}

For Sysbench's degraded reads (Figures~\ref{fig:degraded_reads}(a) and
\ref{fig:degraded_reads}(b)), ZoneAppend-Only suffers severe performance loss,
where \sysname shows 56.3$\times$, 16.5$\times$, 16.5$\times$, and
16.7$\times$ throughput gains over ZoneAppend-Only under a single open segment 
with 4\,KiB chunks and 16\,KiB chunks, as well as multiple open segnents with
$(N_s,N_l)=(2,2)$ and $(N_s,N_l)=(1,3)$, respectively.  The poor performance
of ZoneAppend-Only mainly stems from the high query overhead of the compact
stripe table during data reconstruction.  In all cases, \sysname and
ZoneWrite-Only deliver similar degraded read performance.  

For \texttt{db\_bench}'s degraded reads (Figures~\ref{fig:degraded_reads}(c) and
\ref{fig:degraded_reads}(d)), \sysname achieves 18.5$\times$, 3.5$\times$,
4.3$\times$, and 3.7$\times$ throughput gains over ZoneAppend-Only under the
same four settings, respectively.  Here, compaction is disabled during the
degraded read phase to minimize interference. The performance gap again is due
to ZoneAppend-Only's compact stripe table lookup overhead, which significantly
slows down I/Os in degraded mode.  


\begin{table}[!t]
\footnotesize
\centering
\renewcommand{\arraystretch}{1.15}
\setlength{\tabcolsep}{6pt}
\begin{tabular}{c|c|c|c|c}
\hline
\multirow{2}{*}{ File writes} & Single-4\,KiB & Single-16\,KiB & $(N_s,N_l)=(2,2)$ & $(N_s,N_l)=(1,3)$ \\
\cline{2-5}
& \multicolumn{4}{c}{(ZoneAppend-Only / \bf ZapRAID)} \\
\hline
L2P Table     & 100.00 / \bf 100.00 & 100.00 / \bf 100.00 & 100.00 / \bf 100.00 & 100.00 / \bf 100.00 \\
Segment table & 0.0008 / \bf 0.0008 & 0.0008 / \bf 0.0008 & 0.0008 / \bf 0.0008 & 0.0008 / \bf 0.0008 \\
Compact stripe table           & 106.68 / \bf 35.56  & 26.67 / \bf 8.89    & 26.67 / \bf 0.00    & 26.67 / \bf 0.00 \\
Total     & 206.68 / \bf 135.56 & 126.67 / \bf 108.89 & 126.67 / \bf 100.00 & 126.67 / \bf 100.00 \\
\hline
\multirow{2}{*}{ Key-value store} & Single-4\,KiB & Single-16\,KiB & $(N_s,N_l)=(2,2)$ & $(N_s,N_l)=(1,3)$ \\
\cline{2-5}
& \multicolumn{4}{c}{(ZoneAppend-Only / \bf ZapRAID)} \\
\hline
L2P table     & 77.55 / \bf 77.55 & 77.55 / \bf 77.55 & 77.55 / \bf 77.55 & 77.55 / \bf 77.55 \\
Segment table & 0.0006 / \bf 0.0006 & 0.0006 / \bf 0.0006 & 0.0006 / \bf 0.0006 & 0.0006 / \bf 0.0006 \\
Compact stripe table           & 81.58 / \bf 27.19  & 20.40 / \bf 6.80    & 20.40 / \bf 0.00    & 20.40 / \bf 0.00 \\
Total     & 159.13 / \bf 104.74 & 97.95 / \bf 84.35   & 97.95 / \bf 77.55   & 97.95 / \bf 77.55 \\
\hline
\end{tabular}
\vspace{3pt}
\caption{Exp\#3 (Memory usage for index management). Breakdown of memory usage
(MiB) for index management in ZoneAppend-Only and \sysname under the file
writes and key-value store workloads.}
\label{tab:mem_usage_breakdown}
\end{table}

\para{Exp\#3 (Memory usage for index management).}  To evaluate memory
usage for index management, we consider the file writes and key-value store
workloads.  We assume that the L2P table is entirely in memory, without being
offloaded (\S\ref{subsec:overview}).  Table~\ref{tab:mem_usage_breakdown}
reports the breakdown of memory usage for index management (including the L2P
table, segment table, and compact stripe table) and the total memory usage. 

For file writes, we run Sysbench with a request size of 64\,KiB to generate a
100\,GiB dataset, while for key-value store workloads, we run
\texttt{db\_bench} with compaction enabled to write a total of 100\,GiB of
data across eight clients with 64\,KiB requests.  We compare ZoneAppend-Only
and \sysname, since they both maintain stripe-to-chunk mappings under Zone
Append, while ZoneWrite-Only does not issue Zone Append and hence has no
compact stripe table memory consumption.  The workload settings are identical
to those used in Exp\#1.  

For file writes, the L2P table and segment table sizes stay constant across
configurations (100.00\,MiB and 0.0008\,MiB, respectively), as they depend
only on the total volume of data written.  The differences come from the
compact stripe table.  For a single open segment with 4\,KiB chunks, the
compact stripe table decreases from 106.68\,MiB (ZoneAppend-Only) to
35.56\,MiB (\sysname) (i.e., 66.7\% reduction), so \sysname reduces the total
memory usage of ZoneAppend-Only from 206.68\,MiB to 135.56\,MiB (34.4\%).  We
emphasize that \sysname's memory reduction (66.7\%) of the compact stripe
table compared to ZoneAppend-Only is critical, since it significantly reduces
the degraded read overhead (Exp\#2). 

\sysname also shows memory usage reduction in other settings.  For a single
open segment with 16\,KiB chunks, the compact stripe table drops from
26.67\,MiB to 8.89\,MiB (66.7\%), and \sysname reduces the total memory usage
of ZoneAppend-Only from 126.67\,MiB to 108.89\,MiB (14.0\%).  For multiple
open segments with $(N_s,N_l)=(2,2)$ and $(N_s,N_l)=(1,3)$, \sysname directs
all 64\,KiB writes to large segments, so it reverts to ZoneWrite and does not
require a compact stripe table.  Thus, \sysname reduces the compact stripe
table size from 26.67\,MiB to 0.00\,MiB, and the total memory usage is reduced
from 126.67\,MiB to 100.00\,MiB (21.1\%), compared to ZoneAppend-Only. 
	 
For key-value store workloads, the trends are similar.  The L2P and
segment table sizes remain fixed at 77.55\,MiB and 0.0006\,MiB, respectively,
while the compact stripe table and total memory usage show clear savings in
\sysname.  For example, for a single open segment with 4\,KiB chunks, \sysname
reduces the compact stripe table size from 81.58\,MiB to 27.19\,MiB (66.7\%), 
reducing the total memory usage from 159.13\,MiB to 104.74\,MiB (34.2\%), 
compared to ZoneAppend-Only.

\para{Summary of real-application experiments.}  Compared to
ZoneWrite-Only, \sysname excels in small chunk sizes and increases the write
throughput by up to 76.8\% (in 64\,KiB writes under Sysbench).  A trade-off is
that \sysname introduces a compact stripe table, but it limits the memory
usage of the compact stripe table (as well as the degraded read overhead)
through the group-based data layout. 
	
Compared to ZoneAppend-Only, \sysname achieves nearly identical write
throughput (Exp\#1), while significantly improving degraded read performance
(Exp\#2) and reducing memory usage (Exp\#3).  For example, in Sysbench file
writes under a single open segment with 4\,KiB chunks, \sysname's write
throughput is within 2.42\% of ZoneAppend-Only's, while reducing the compact
stripe table size by 66.7\% and the total memory usage for index management by
34.4\%.  Similarly, in the \texttt{db\_bench} \texttt{fillrandom} workload
under a single segment with 4\,KiB chunks, \sysname's write throughput is
within 2.20\% of ZoneAppend-Only's, while reducing the compact stripe table
size by 66.7\% and the total memory usage for index management by 34.2\%.  By
limiting the compact stripe table size, \sysname also significantly
outperforms ZoneAppend-Only in degraded reads.

\subsection{Trace-Driven Experiments (RQ2)}
\label{subsec:trace}

We evaluate via trace-driven evaluation \sysname on real-world cloud block
storage workloads using the block I/O traces from Alibaba Cloud \cite{li20}.
We configure the logical space capacity as 500\,GiB.  Our evaluation considers
four settings of open segments: (i) a single open segment with a chunk size of
4\,KiB, (ii) a single open segment with a chunk size of 16\,KiB, (iii) four
open segments with $(N_s, N_l) = (2,2)$, and (iv) four open segments with
$(N_s, N_l) = (1,3)$; for (iii) and (iv), we set $C_s=$~8\,KiB and
$C_l=$~16\,KiB. 

\para{Exp\#4 (Cloud block storage workloads).} The original Alibaba
Cloud traces cover 1,000 volumes \cite{li20}. We identify that 684 out of the
1,000 volumes have more than 60\% of small write requests of no more than
4\,KiB and more than 25\% of large write requests of at least 16\,KiB
(\S\ref{subsec:vs}).  Among the 684 volumes, we choose 20 of them whose
amounts of write traffic range between 200\,GiB and 240\,GiB.  Note that the
selected 20 volumes are dominated by writes: in 18 out of 20 volumes, more
than 88.3\% of requests are writes, while in all volumes, more than 67.0\% of
requests are writes.  We conduct our evaluation over five runs; in each run,
we evaluate the average throughput and 95th-percentile latency of all read and
write operations when running each scheme over the 20 selected volumes. 

\begin{figure}[!t]
\centering
\begin{tabular}{@{\ }c@{\ }c@{\ }c@{\ }c}
\multicolumn{4}{c}{
\includegraphics[width=3.6in]{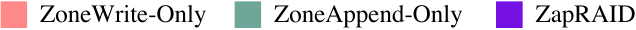}}\\
\includegraphics[width=1.46in]{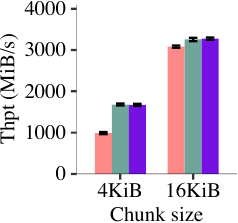} &
\includegraphics[width=1.46in]{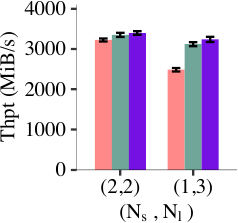} &
\includegraphics[width=1.46in]{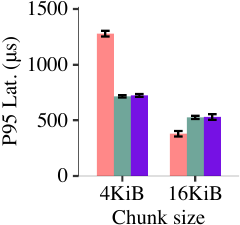} &
\includegraphics[width=1.46in]{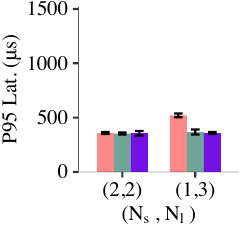} \\ 
\makecell{\small (a) Throughput \\ \small (single segment)} &
\makecell{\small (b) Throughput \\ \small (multiple segments)} &
\makecell{\small (c) p95 latency \\ \small (single segment)}&
\makecell{\small (d) p95 latency \\ \small (multiple segments)}
\end{tabular}
\vspace{-6pt}
\caption{Exp\#4 (Cloud block storage workloads). Average throughput and 95
latency results of all 20 volumes in Alibaba Cloud traces.}
\label{fig:trace}
\end{figure}

Figures~\ref{fig:trace}(a) and \ref{fig:trace}(b) show the throughput results. 
For a single open segment with 4\,KiB and 16\,KiB chunks
(Figure~\ref{fig:trace}(a)), the throughput of \sysname is 69.4\% and 6.4\%
higher than that of ZoneWrite-Only, respectively, and is close to that of
ZoneAppend-Only;  For multiple open segments $(N_s, N_l) = (2,2)$, all
three schemes have similar throughput, while for $(N_s, N_l) = (1,3)$,
\sysname has 25.3\% higher throughput than ZoneWrite-Only and similar
throughput to ZoneAppend-Only. 

Figures~\ref{fig:trace}(c) and \ref{fig:trace}(d) show the 95th-percentile
latencies. For a single open segment with 4\,KiB and 16\,KiB chunks
(Figure~\ref{fig:trace}(c)), \sysname has 56.6\% lower and 39.6\% higher
95th-percentile latencies than ZoneWrite-Only under the chunk sizes of 4\,KiB
and 16\,KiB, respectively, and comparable 95th-percentile latencies with
ZoneAppend-Only. For multiple open segments (Figure~\ref{fig:trace}(d)), all
three schemes have similar latencies for $(N_s, N_l) = (2, 2)$, while
ZoneWrite-Only has a higher 95th-percentile latency for $(N_s, N_l) = (1,3)$ as
its performance is bottlenecked by the small writes that are forwarded to the
only small-chunk open segment. 

\begin{table}[!t]
\footnotesize
\centering
\renewcommand{\arraystretch}{1.15}
\setlength\tabcolsep{2.5pt}
\begin{tabular}{c|c|c|c|c|c|c|c|c|c|c|c}
\hline
\multicolumn{2}{c|}{\bf Volume No.} & 
    {\bf 1} & {\bf 2} & {\bf 3} & {\bf 4} & {\bf 5} &
    {\bf 6} & {\bf 7} & {\bf 8} & {\bf 9} & {\bf 10} \\ 
\hline
\multirow{2}{*}{Ratios (\%)} & 
Small writes & 
    83.35 & 83.06 & 83.30 & 83.56 & 83.29 &
    83.50 & 83.03 & 83.36 & 82.66 & 82.98 \\ 
\cline{2-12}
& Large writes & 
    1.63 & 1.64 & 1.67 & 1.70 & 1.75 &
    1.76 & 1.77 & 1.78 & 1.83 & 1.86 \\
\hline
\multirow{2}{*}{\makecell{Thpt (MiB/s) \\ (4\,KiB chunks)}} &
ZoneWrite-Only &
    1,023.7 & 947.0   & 1,035.1 & 1,001.5 & 937.5 &
    981.5   & 1,035.3 & 968.2   & 986.2   & 988.7 \\ 
\cline{2-12}
& \sysname &
    1,740.6 & 1,711.3 & 1,702.4 & 1,699.1 & 1,660.1 &
    1,657.9 & 1,612.7 & 1,619.4 & 1,681.0 & 1,636.3 \\ 
\hline
\multirow{2}{*}{\makecell{Thpt (MiB/s) \\ (($N_s, N_l$)=(1,3))}} &
ZoneWrite-Only &
    2,486.4 & 2,341.2 & 2,323.1 & 2,401.3 & 2,403.7 &
    2,333.7 & 2,311.0 & 2,360.3 & 2,344.8 & 2,410.5 \\ 
\cline{2-12}
& \sysname &
    3,175.9 & 3,225.1 & 3,271.9 & 3,193.7 & 3,183.1 &
    3,227.5 & 3,231.6 & 3,205.8 & 3,249.7 & 3,298.2 \\ 
\hline
\multicolumn{2}{c|}{\bf Volume No.} & 
    {\bf 11} & {\bf 12} & {\bf 13} & {\bf 14} & {\bf 15} &
    {\bf 16} & {\bf 17} & {\bf 18} & {\bf 19} & {\bf 20} \\ 
\hline
\multirow{2}{*}{Ratios (\%)} & 
Small writes & 
    82.96 & 86.68 & 84.00 & 81.58 & 80.98 & 
    81.41 & 72.45 & 73.85 & 65.78 & 63.05 \\ 
\cline{2-12}
& Large writes & 
    1.92  & 1.95  & 3.42  & 3.70  & 4.49 &
    10.33 & 16.83 & 20.95 & 21.11 & 24.89 \\
\hline
\multirow{2}{*}{\makecell{Thpt (MiB/s) \\ (4\,KiB chunks)}} &
ZoneWrite-Only &
    948.4   & 1,007.2 &  982.2  &  995.6   &  989.2 & 
    972.1   & 970.4   &  993.0  & 1,004.0  &  971.5 \\ 
\cline{2-12}
& \sysname &
    1,627.3 & 1,710.9 & 1,634.0 & 1,645.8 & 1,699.1 &
    1,666.3 & 1,684.8 & 1,736.4 & 1,656.4 & 1,665.6 \\ 
\hline
\multirow{2}{*}{\makecell{Thpt (MiB/s) \\ (($N_s, N_l$)=(1,3))}} &
ZoneWrite-Only &
    2,372.7 & 2,399.1 & 2,412.2 & 2,439.7 & 2,593.6 &
    2,637.1 & 2,703.2 & 2,773.4 & 2,862.1 & 2,825.4 \\
\cline{2-12}
& \sysname &
    3,175.9 & 3,225.1 & 3,271.9 & 3,193.7 & 3,183.1 &
    3,227.5 & 3,231.6 & 3,205.8 & 3,249.7 & 3,298.2 \\ 
\hline
\end{tabular}
\vspace{3pt}
\caption{Exp\#4 (Cloud block storage workloads).  Statistics of the
20 individual volumes in Alibaba Cloud traces.}
\label{tab:trace}
\vspace{-9pt}
\end{table}

We also show the statistics of individual volumes by presenting the
ratios of numbers of small writes (i.e., 4\,KiB) and large writes (i.e., at
least 16\,KiB) over the total number of write requests as well as the
throughput of each volume.  For brevity, since all three schemes have
close throughput under one single open segment with 16\,KiB chunks and four
open segments with $(N_s, N_l)=(2,2)$, we only present the results of one
single open segment with 4\,KiB chunks and four open segments with $(N_s,
N_l)=(1,3)$.  Also, we only focus on the results under ZoneWrite-Only and
\sysname to show their variations across the volumes. 

Table~\ref{tab:trace} shows the results. We sort the volumes by the ratio of
large write requests. For one single open segments under 4\,KiB chunks, the
throughput is close over all volumes for each scheme. For four open segments
with $(N_s, N_l)=(1,3)$, ZoneWrite-Only shows lower throughput for a lower
ratio of large writes; for example, the throughput results of Volumes 2 and 20
are 2,341.2\,MiB/s and 2,825.4\,MiB/s with large-write ratios of 1.64\% and
24.89\%, respectively. The reason is that with fewer large writes, more write
requests are written to the only $N_s=1$ small-chunk open segment, and
ZoneWrite-Only fails to utilize intra-zone parallelism to issue writes
efficiently. In contrast, \sysname maintains higher throughput with an
increase of 14.7-40.8\% compared with ZoneWrite-Only under $(N_s, N_l)=(1,3)$,
across all 20 volumes.

\subsection{Microbenchmarks (RQ3)}
\label{subsec:micro}

We examine the performance of \sysname in writes, normal reads, degraded
reads, crash recovery, and full-drive recovery.  In our default setup, we
configure the logical space capacity as 200\,GiB and set $G=256$ for \sysname
in its group-based data layout.  First, we focus on one open segment
with various chunk sizes in Exp\#5-Exp\#10.  We do not consider garbage
collection (by limiting the total write size to be less than the logical space
capacity) and keep the L2P table entirely in memory. We also consider
the impact of various group sizes (Exp\#7), the impact of RAID schemes
(Exp\#8), as well as the crash recovery and full-drive recovery
performance with different logical space capacities (Exp\#9).
Furthermore, we consider multiple open segments in Exp\#11-Exp\#13,
including the write performance under multiple open segments
(Exp\#11), the overhead of garbage collection (Exp\#12), and
the overhead of L2P table offloading (Exp\#13). 

\para{Exp\#5 (Write performance).} We first
evaluate the write performance of \sysname on a single open segment.  We use
Flexible IO Tester (FIO) (v3.35) \cite{fio} to generate a write-only workload
that issues random writes of 64\,GiB of data.  We set the queue depth as 64 to
saturate the system parallelism (Exp\#10).  We compare \sysname with
RAIZN-SPDK, ZoneWrite-Only, and ZoneAppend-Only.

\begin{figure}[!t]
\centering
\begin{tabular}{@{\ }c@{\ }c}
\multicolumn{2}{c}{
\includegraphics[width=3.2in]{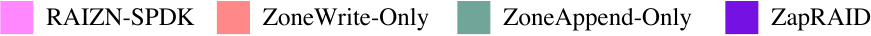}} \\
\includegraphics[width=2.7in]{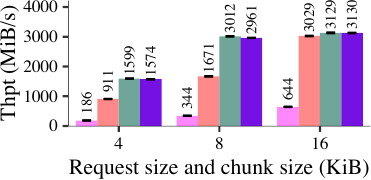} & 
\includegraphics[width=2.7in]{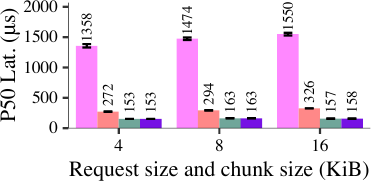} \\ 
\makecell[c]{\small (a) Write throughput} &
\makecell[c]{\small (b) Median write latency} 
\vspace{6pt}\\ 
\includegraphics[width=2.7in]{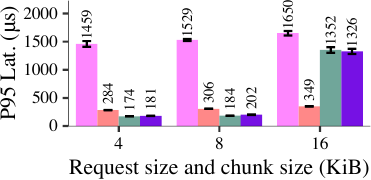} & 
\includegraphics[width=2.7in]{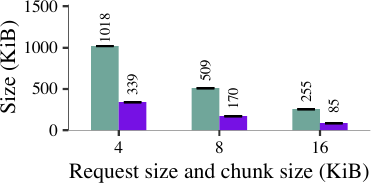} \\ 
\makecell[tc]{\small (c) p95 write latency} & 
\makecell[tc]{\small (d) Memory usage (per GiB of data)\\ 
			\small for compact stripe table}
\end{tabular}
\vspace{-6pt}
\caption{Exp\#5 (Write performance on a single open segment). We show
the results when the request size is equal to the chunk size. For
figure~(d), we show ZoneAppend-Only and \sysname only.}
\label{fig:write}
\end{figure}

First, we issue write requests of 4\,KiB, 8\,KiB, and 16\,KiB, where the
request size is the same as the chunk size.
Figures~\ref{fig:write}(a)-\ref{fig:write}(c) show the throughput as well as
the median and 95th-percentile latencies. Compared with ZoneWrite-Only, for
4\,KiB and 8\,KiB writes, \sysname achieves 72.8\% and
77.2\% higher write throughput, 43.8\% and 44.5\% lower median latency, and
36.4\% and 34.1\% lower 95th-percentile latency, respectively.  This shows
that \sysname improves the write performance by exploiting the intra-zone
parallelism using Zone Append for small writes.  On the other hand, for 16\,KiB
writes, \sysname has similar write throughput and median latency, but its
95th-percentile latency is $3.80\times$ compared with that of ZoneWrite-Only.
A possible reason is that Zone Append has high computational overhead in the
ZNS SSD firmware, which incurs higher delays in processing individual Zone
Append commands. 

Note that in Figure~\ref{fig:write}(a), the throughput of 4\,KiB writes
under ZoneWrite-Only (910.8\,MiB/s) is higher than that reported in our
conference paper \cite{wang23} (662.4\,MiB/s).  The reason is that we now
optimize the implementation of ZoneWrite-Only by parallelizing the RAID
operations (e.g., parity computations and stripe metadata updates) of the
current stripe and the writes of the next stripe (as opposed to serially
executing the operations).  Now, the throughput is close to 3$\times$ the
write throughput of Zone Write to an open zone (337.6\,MiB/s in
\S\ref{subsec:zns}).  This implies that the main bottleneck of ZoneWrite-Only
lies in I/Os rather than computations. 

Compared with ZoneAppend-Only, \sysname achieves similar throughput and
latencies for all write sizes, even though \sysname needs to wait for the
completion of all Zone Append commands in a stripe group before issuing the
writes for the next stripe group.  Note that \sysname has much less memory
usage (\S\ref{subsec:group}) and much lower degraded read latency (Exp\#7)
than ZoneAppend-Only. 

Compared with RAIZN-SPDK, all other three schemes show much higher performance
on a single segment. The reasons are two-fold. First, RAIZN does not support
Zone Append for intra-zone parallelism. Second, it uses partial parity updates
to acknowledge individual write requests instead of issuing full-stripe writes
as in the other three schemes.  This limits the concurrency of write requests
and hence the overall write performance. In Exp\#11, we further compare
\sysname with RAIZN-SPDK under multiple open segments.

Figure~\ref{fig:write}(d) shows the memory usage for stripe management
per GiB of written data. Here, we focus on ZoneAppend-Only and \sysname, as
they both need to maintain mappings between stripes and chunk locations under
Zone Append (note that ZoneWrite-Only has no compact stripe table as it does
not issue Zone Append).  For 4\,KiB writes, ZoneAppend-Only consumes
1,018.2\,KiB of memory, while \sysname consumes 339.4\,KiB only (i.e., 66.7\%
reduction) due to its compact stripe table design.  For a (3+1)-RAID-5 array
in which each drive has $Z=$~3,690 zones and the zone capacity is 1,077\,MiB,
the maximum memory size for ZoneAppend-Only is $\frac{1018.2 \times 3 \times
3690\times 1077}{1024 \times 1024 \times 1024} = 11.3$\,GiB, while \sysname
consumes 3.77\,GiB only (\S\ref{subsec:group}). The result suggests that
\sysname greatly reduces the memory footprints for stripe management compared
with ZoneAppend-Only.  Note that larger chunks imply a smaller compact stripe
table since each segment has fewer chunks. 

\begin{figure}[!t]
\centering
\begin{tabular}{@{\ }c@{\ }c}
\multicolumn{2}{c}{
\includegraphics[width=3.2in]{figs/exp1.singlezone/pdf/exp1_write_legend.pdf}} \\
\includegraphics[width=2.7in]{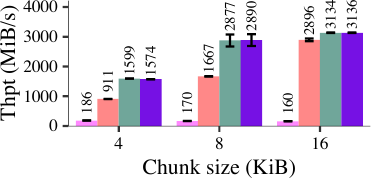} & 
\includegraphics[width=2.7in]{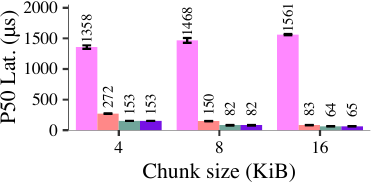} \\ 
\makecell[c]{\small (a) Write throughput} &
\makecell[c]{\small (b) Median write latency} 
\vspace{6pt}\\ 
\includegraphics[width=2.7in]{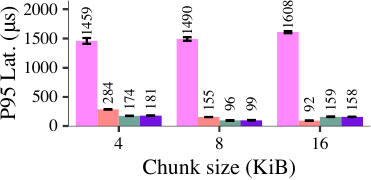} & 
\includegraphics[width=2.7in]{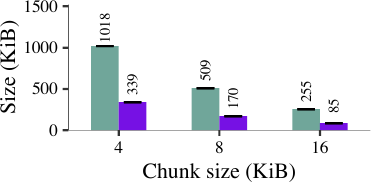} \\
\makecell[tc]{\small (c) p95 write latency} & 
\makecell[tc]{\small (d) Memory usage (per GiB of data)\\ 
			\small for compact stripe table}
\end{tabular}
\vspace{-6pt}
\caption{Exp\#5 (Write performance on a single open segment). We show
the results when the request size is fixed as 4\,KiB.}
\label{fig:write_4kreq}
\end{figure}

We also issue write requests of a fixed size of 4\,KiB, while varying the
chunk size, to show the impact of the chunk size on write performance.
Figure~\ref{fig:write_4kreq} shows the results. \sysname still
maintains the high throughput and low median latencies for various chunk sizes,
but its 95th-percentile latency for 16\,KiB chunks is 50.7\% higher than that
for 8\,KiB chunks.  Similar results also appear in ZoneAppend-Only.  The result
indicates that Zone Append is mainly beneficial for small writes. It also
justifies why we choose Zone Write over Zone Append in hybrid data management
(\S\ref{subsec:hybrid}). 

\para{Exp\#6 (Normal and degraded read performance).} We compare the
performance of both normal reads and degraded reads in Log-RAID and \sysname. 
We vary the chunk size and set the read size to be the same as the chunk size.
For each chunk size, we first fill up the 200\,GiB logical address space to
ensure that every logical block being read is indeed physically stored.  We
then use FIO to generate a read-only workload that issues random reads to
16\,GiB of data using the {\tt randread} option in FIO.  We set the queue
depth as one, so as to focus on the performance of individual read requests
and exclude the interference among requests.  We consider three cases: normal
reads (NR), degraded reads under the static mapping in Log-RAID (DR-Log-RAID)
(\S\ref{subsec:lograid}), and degraded reads under the group-based data layout
in \sysname (DR-ZapRAID). Note that both Log-RAID and \sysname have the same
workflow for normal reads.  To evaluate degraded reads, we fail a drive and
issue reads to the lost blocks of the failed drive. 

\begin{figure}[!t]
\centering
\begin{tabular}{@{\ }c@{\ }c@{\ }c}
\multicolumn{3}{c}{
\includegraphics[width=2.0in]{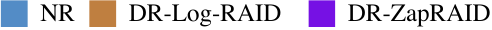}} \\
\includegraphics[width=2in]{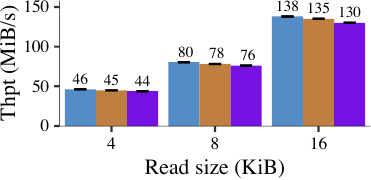} & 
\includegraphics[width=2in]{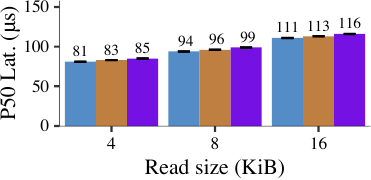} & 
\includegraphics[width=2in]{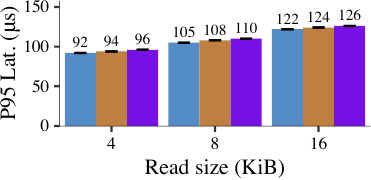} \\
{\small (a) Read throughput}&
{\small (b) Median read latency}&
{\small (c) p95 read latency}
\end{tabular}
\vspace{-6pt}
\caption{Exp\#6 (Normal and degraded read performance). We consider normal
reads (NR), degraded reads under static mapping in Log-RAID (DR-Log-RAID), and
degraded reads under the group-based data layout in \sysname (DR-\sysname).} 
\label{fig:read}
\end{figure}

Figure~\ref{fig:read} shows the throughput as well as the median and
95th-percentile latencies for different read sizes (i.e., chunk sizes).  All
three read operations have less than 6\% difference. The degraded reads of
both Log-RAID and \sysname have slightly worse performance than the normal
reads, since both systems retrieve multiple available chunks in parallel for
decoding.  Also, \sysname achieves comparable performance to Log-RAID, which
uses static mapping. 

\para{Exp\#7 (Impact of stripe group size).} We study the impact of the
stripe group size $G$ on the write and read performance in \sysname. We vary
$G$ from 4 to 4,096. For both writes and reads, we vary the chunk size and
set the request size to be the same as the chunk size.

\begin{figure}[!t]
\centering
\begin{tabular}{@{\ }cc}
\multicolumn{2}{c}{
\includegraphics[width=4.2in]{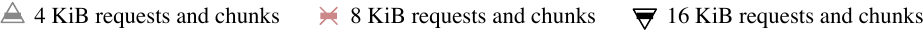}} \\
\multicolumn{2}{c}{
\includegraphics[width=2.6in]{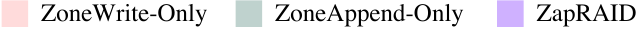}} \\
\includegraphics[width=3in]{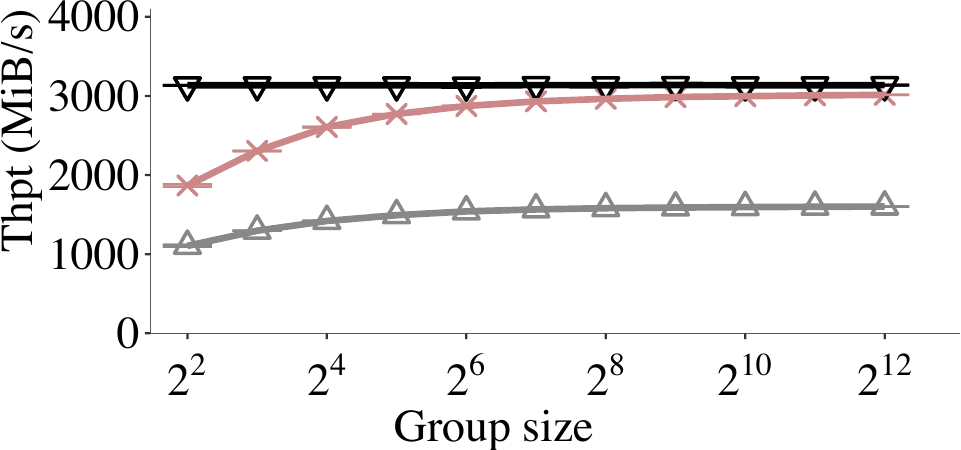} & 
\includegraphics[width=3in]{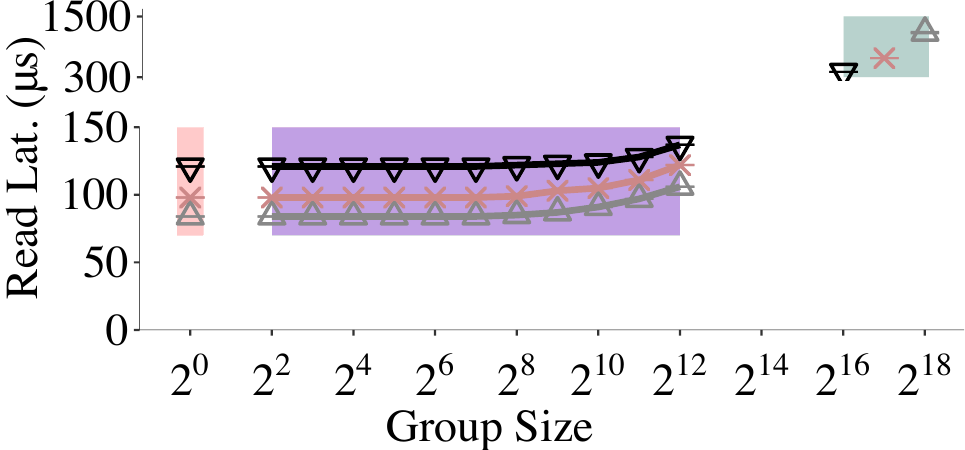}\\
\makecell[tc]{\small (a) Write throughput for \sysname}&
\makecell[tc]{\small (b) Median degraded read latencies\\
  	\small for different schemes}
\end{tabular}
\vspace{-6pt}
\caption{Exp\#7 (Impact of stripe group size).} 
\label{fig:sen}
\end{figure}

Figure~\ref{fig:sen} shows the write throughput and the median degraded read
latency of \sysname versus $G$.  The write throughput of \sysname increases
with $G$ for $G\le 256$ and remains stable for $G\ge 256$ under 4\,KiB and
8\,KiB requests, while it remains almost unchanged under 16\,KiB requests.  The
write throughput reaches 1,574.2\,MiB/s and 2,961.2\,MiB/s for $G=256$ under
the chunk size of 4\,KiB and 8\,KiB (1.43$\times$ and 1.59$\times$ the write
throughput for $G=4$), respectively. Thus, \sysname can better exploit the
intra-zone parallelism through Zone Append as $G$ increases, but the
parallelism saturates when $G$ becomes sufficiently large. For 16\,KiB chunks,
the intra-zone parallelism brought by Zone Append is limited and already
saturates when $G$ is small.  For the degraded read latency, when $G$ is
small, the query overhead of the compact stripe table is negligible (around
1\,$\mu$s based on our measurement).  However, as $G$ continues to increase,
the query overhead becomes non-negligible. For example, the degraded read
latency for $G=$~4,096 increases by 13.2-24.7\% compared with that for
$G=256$.  In addition to I/O performance, a large $G$ also increases the
memory usage of the compact stripe table (\S\ref{subsec:group}). Thus, we
choose $G=256$ as our default setting. 

Figure~\ref{fig:sen}(b) presents the median degraded read latencies of
ZoneWrite-Only, ZoneAppend-Only. Note
that $G=1$ represents ZoneWrite-Only, while $G=S$ represents ZoneAppend-Only. 
We use different colored rectangles to differentiate the three schemes.  The
parameter $S$ varies with the chunk size, as larger chunks imply fewer chunks
per zone (i.e., a smaller $S$).  The median degraded read latency of
ZoneAppend-Only is significantly higher than that of \sysname for $G=256$
across different chunk sizes (13.9$\times$, 6.82$\times$, and 3.31$\times$
under 4\,KiB, 8\,KiB, and 16\,KiB chunks, respectively).
Thus, ZoneAppend-Only not only incurs high memory usage (\S\ref{subsec:group}
and Exp\#5), but also has poor degraded read performance due to its
high query overhead for the mappings between stripes and chunk locations. 

\para{Exp\#8 (Impact of RAID schemes).}  \sysname supports various RAID
schemes. We configure RAID-0, RAID-01, RAID-4, RAID-5, and RAID-6 on the four
ZNS SSDs (e.g., $k=2$ and $m=2$ in RAID-6).  Here, we compare \sysname and
ZoneWrite-Only. We fix the request size to be the same as the chunk size and
vary the chunk size with a queue depth of 64. 

\begin{figure}[!t]
\centering
\begin{tabular}{@{\ }c@{\ }c@{\ }c}
\multicolumn{3}{c}{
\includegraphics[width=1.6in]{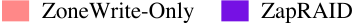}} \\ 
\includegraphics[width=2in]{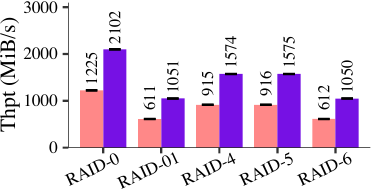} &
\includegraphics[width=2in]{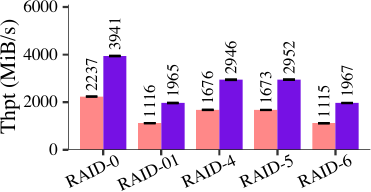} & 
\includegraphics[width=2in]{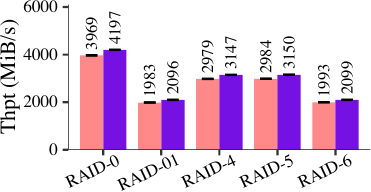} \\
{\small (a) 4\,KiB chunks} &
{\small (b) 8\,KiB chunks} &
{\small (c) 16\,KiB chunks}
\end{tabular}
\vspace{-6pt}
\caption{Exp\#8 (Impact of RAID schemes).} 
\label{fig:raidscheme}
\end{figure}

Figure~\ref{fig:raidscheme} depicts the results. \sysname increases the write
throughput of ZoneWrite-Only by 71.5-72.1\%, 75.8-76.4\%, and 5.3-5.7\% for
4\,KiB, 8\,KiB, and 16\,KiB chunks, respectively. The difference in the write
throughput among the RAID schemes is due to the different number of data
chunks in each stripe (i.e., four, two, three, three, and two for RAID-0,
RAID-01, RAID-4, RAID-5, and RAID-6, respectively). The results show that
\sysname consistently increases the write throughput of ZoneWrite-Only under
different RAID schemes.

\para{Exp\#9 (Recovery performance).} 
We evaluate the recovery performance of \sysname in two aspects: recovery from
a system crash (\S\ref{subsec:crash}) and recovery from a full-drive failure
(\S\ref{subsec:all}).  We sequentially write data to \sysname configured with
a fixed size of logical space (varying from 100\,GiB to 1,000\,GiB), mimic
each type of failure, and repeat the recovery procedure for five runs.

\begin{figure}[!t]
\centering
\begin{tabular}{@{\ }cc}
\includegraphics[width=2.7in]{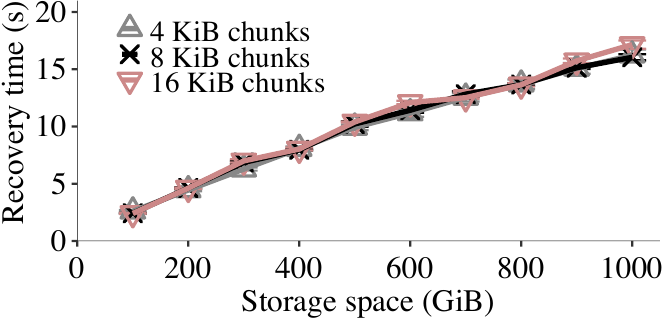}&
\includegraphics[width=2.7in]{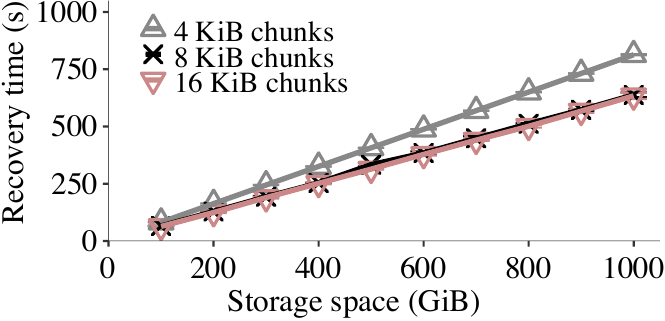}\\
{\small (a) Crash recovery}&
{\small (b) Full-drive recovery}
\end{tabular}
\vspace{-6pt}
\caption{Exp\#9 (Recovery performance).}
\label{fig:recovery}
\end{figure}

We first evaluate the crash recovery time of \sysname.  We crash \sysname and
rebuild it from the currently stored data in the drives.  We report the
average recovery time (i.e., the time spent on performing all recovery
operations in \S\ref{subsec:crash}).  Figure~\ref{fig:recovery}(a) depicts the
results for different chunk sizes. For the 100\,GiB storage space, \sysname
takes 2.38-2.60\,s to recover the system state. The recovery time increases
linearly with the logical space size.  The reason is that for larger logical
space, \sysname needs to recover the system state from more sealed segments by
reading their footer regions.  For example, for the 1,000\,GiB storage space,
the recovery time increases to 15.92-16.22\,s.  Nevertheless, the recovery
overhead is limited.  For example, for 4\,KiB chunks, we only need to pay
an additional $(16.22 - 2.60) / (10 - 1) = 1.51$\,s for recovering the system
state from each 100\,GiB of additional logical space.  Also, the crash
recovery time remains almost the same for different chunk sizes, since most of
the time is spent on reading the footer regions with a large read size,
independent of the chunk size. 

We next evaluate the full-drive recovery time of \sysname.  We erase all data
in one drive and recover the lost data in the same drive.  We report the
average recovery time as the total time spent on reading available data from
existing drives, reconstructing the stripes, and writing the recovered data to
the new drive. Figure~\ref{fig:recovery}(b) shows the results for different
chunk sizes. The full-drive recovery time is proportional to the logical space
size. For example, for 4\,KiB chunks, \sysname takes 81.9\,s and 813.2\,s to
recover the lost data for the 100\,GiB and 1,000\,GiB storage space,
respectively. On average, for each 100\,GiB storage space, \sysname spends
$813.2 / 10 = 81.3$\,s to recover the lost data.  Note that the recovery time
is smaller for larger chunk sizes since \sysname can write recovered data with
a larger write size. For example, for 8\,KiB and 16\,KiB chunks, the recovery
time is reduced by 18.1-22.2\% and 22.1-23.8\% compared with 4\,KiB chunks,
respectively.

\para{Exp\#10 (Scalability).} We examine the scalability of \sysname by
evaluating its write performance under different values of queue depth and
showing how it saturates the system parallelism with a high queue depth.  We
consider both real and FEMU-emulated \cite{li18} ZNS SSDs. 

We first vary the queue depth in our testbed with the real ZNS SSDs.  We use
the write-only workload, vary the chunk size, and set the write size to be the
same as the chunk size.  We set the minimum queue depth as four to evaluate
\sysname under concurrent writes. 

\begin{figure}[!t]
\centering
\begin{tabular}{@{\ }ccc}
\multicolumn{3}{c}{
\includegraphics[width=4.2in]{figs/exp3.groupSize/pdf/exp3_legend.pdf}} \\
\includegraphics[width=2in]{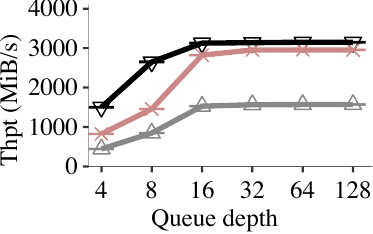}&
\includegraphics[width=2in]{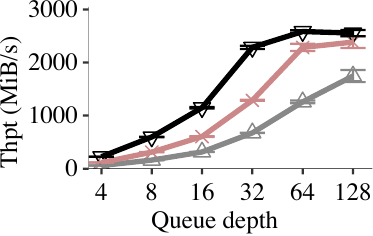}&
\includegraphics[width=2in]{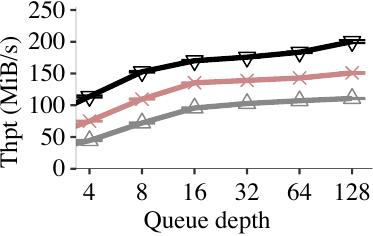}\\
\makecell{\small (a) \sysname on real ZNS SSDs}&
\makecell{\small (b) \sysname on FEMU ZNS SSDs} &
\makecell{\small (c) Conventional RAID on FEMU\\ block-interface SSDs}
\end{tabular}
\vspace{-6pt}
\caption{Exp\#10 (Scalability). The results in figures~(a) and (b) are
based on \sysname, and the results in figure~(c) are based on the conventional
RAID array.}
\label{fig:scale}
\end{figure}

Figure~\ref{fig:scale}(a) depicts the results.  \sysname increases the write
throughput when the queue depth increases. It saturates the system parallelism
when the queue depth is 16, in which the throughput gain is 3.52$\times$,
3.57$\times$, and 2.08$\times$ compared with when the queue depth is four
for 4\,KiB, 8\,KiB, and 16\,KiB chunks, respectively. Note that
we set the queue depth for all microbenchmark experiments as 64 to further
increase the throughput for multiple open segments (Exp\#11). 

Our current testbed has limited parallelism as it only has four real ZNS SSDs.
To examine the scalability of \sysname with a higher degree of parallelism, we
have also built an emulated ZNS SSD array with FEMU \cite{li18}.  We set up a
(6+1)-RAID-5 array on seven FEMU SSDs, using 4$\times$128\,GiB Intel Optane
Persistent Memory Modules \cite{optane} in Memory Mode as the emulated
persistent storage backend. Each FEMU SSD is configured with four channels
with four chips each.  Each chip has 64 flash blocks with 4,096 4\,KiB flash
pages each.  Each zone spans 16 chips, meaning that I/Os can be issued to up
to 16 chips in parallel.  We configure the page write latency as 140\,$\mu$s
\cite{li21} and do not emulate write buffers in SSDs that can reduce write
latency.  The page write latency translates to the write throughput of
27.9\,MiB/s per chip, or equivalently, 446.4\,MiB/s per zone if the
parallelism is fully exploited. Since the FEMU SSD does not support
flash-level page metadata, we store the page metadata in an in-memory array.
We configure seven FEMU SSDs in a FEMU virtual machine, installed with Ubuntu
22.04, 32\,GiB memory, and 12~vCPUs. We export the \sysname prototype as a
block device with 96\,GiB of logical space.  We use FIO to issue the random
writes of 64\,GiB of data with a block size of 4\,KiB to \sysname in the
virtual machine for various chunk sizes and queue depths. 

Figure~\ref{fig:scale}(b) shows the results.  When the queue depth is four,
the write throughput for 4\,KiB, 8\,KiB, and 16\,KiB chunks are 58.3\,MiB/s,
114.0\,MiB/s, and 224.2\,MiB/s, respectively.  As the queue depth increases,
the write throughput of \sysname increases and achieves 1,745.5\,MiB/s,
2,384.6\,MiB/s, and 2,554.2\,MiB/s (i.e., 29.9$\times$, 20.9$\times$, and
11.4$\times$ compared with the queue depth of four), respectively when the
queue depth is 128.  

We also examine the performance of the conventional RAID array based
on FEMU-emulated block-interface SSDs. Similar to FEMU ZNS SSDs, we configure
a (6+1)-RAID-5 array on seven FEMU black-box SSDs with Linux {\tt mdadm}
\cite{mdadm}. The configurations of flash blocks, channels, and chips are the
same as FEMU ZNS SSDs, and each FEMU SSD has 16\,GiB of physical space. We
also keep the same page write latency and the same settings as the FEMU virtual
machine.  We export the RAID array as a block device with 64\,GiB of logical
space, so the overprovisioned space is about $(\frac{16\times 6}{64}-1) \times
100\%=50\%$. We use FIO to issue random writes of 64\,GiB to the RAID array
for various chunk sizes and queue depths. Figure~\ref{fig:scale}(c) shows the
results. Compared with \sysname in Figure~\ref{fig:scale}(b), the throughput
of the conventional RAID array is much lower (up to 200\,MiB/s only). The
results are expected, since updating a data chunk in the conventional RAID
array issues a read-modify-write operation, which reads the current data chunk
being modified and the current parity chunk, modifies the parity chunk, and
writes back the new data chunk and new parity chunk.  In contrast, \sysname
issues concurrent write requests to form full stripes and avoids the overhead
of parity updates.

\para{Exp\#11 (Multiple open segments).} We evaluate the write performance
\sysname on multiple open segments based on hybrid data management
(\S\ref{subsec:hybrid}) by varying the numbers of small-chunk and large-chunk
segments (i.e., $N_s$ and $N_l$, respectively).  As introduced in
\S\ref{subsec:app}, we fix $C_s=$\,8\,KiB,
$C_l=$\,16\,KiB, and the number of open segments as four.  We consider
different combinations of $N_s$ and $N_l$, such that $N_s + N_l = 4$.  Note
that if $N_s = 0$ (resp. $N_l=0$), all writes are issued to large-chunk (resp.
small-chunk) segments. 

We consider four workloads. The first three workloads issue 4\,KiB, 8\,KiB, and
16\,KiB writes via FIO, respectively.  Note that these workloads do not use all
open segments since they write 4\,KiB or 8\,KiB requests to small-chunk segments
only and 16\,KiB requests to large-chunk segments only.  The fourth workload
consists of a mixture of 75\% 4\,KiB writes and 25\% 16\,KiB writes by setting
the {\tt bssplit} parameter in FIO, based on the observation that 25\% of
write requests in real-world cloud block storage workloads are at least
16\,KiB and most of the write requests are smaller than 4\,KiB \cite{li20}.  
We set the total write size of each workload as 64\,GiB.

\begin{figure}[!t]
\centering
\begin{tabular}{@{\ }cc}
\multicolumn{2}{c}{
\includegraphics[width=2.6in]{figs/exp7.multizone/pdf/exp2_write_legend.pdf}}\\
\includegraphics[width=3in]{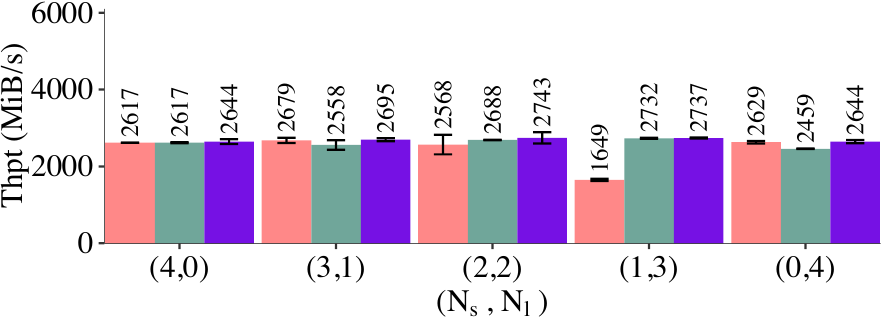} & 
\includegraphics[width=3in]{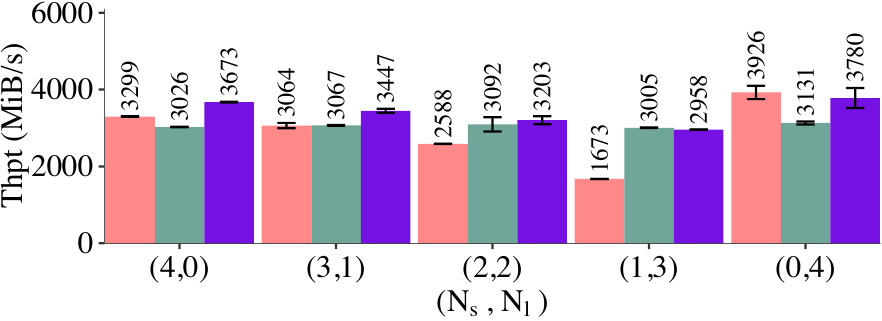} \\ 
{\small (a) 4\,KiB writes, throughput} &
{\small (b) 8\,KiB writes, throughput} 
\vspace{6pt}\\
\includegraphics[width=3in]{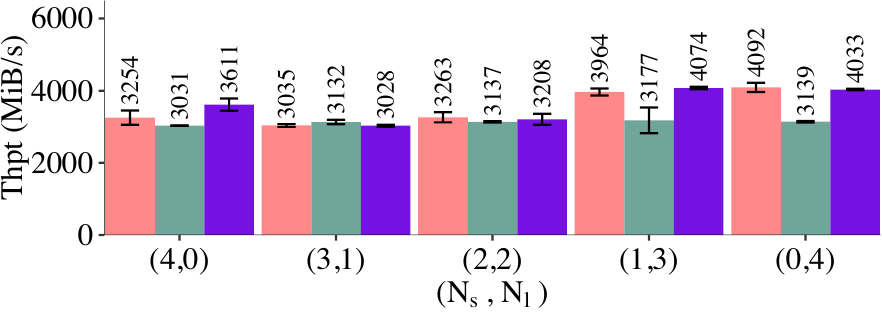} & 
\includegraphics[width=3in]{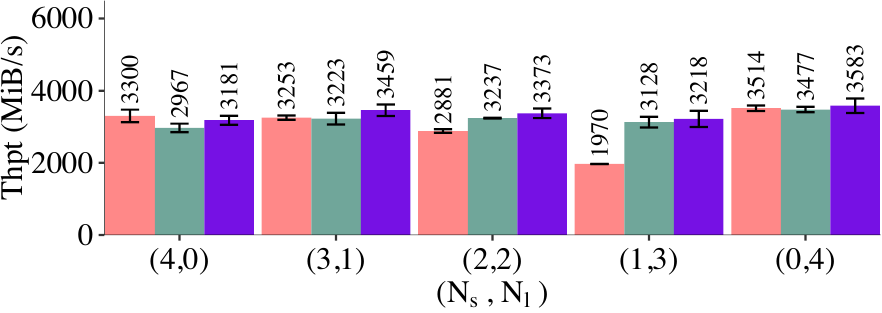} \\
{\small (c) 16\,KiB writes, throughput} &
{\small (d) Hybrid 4\,KiB and 16\,KiB writes, throughput} 
\vspace{6pt}\\
\includegraphics[width=3in]{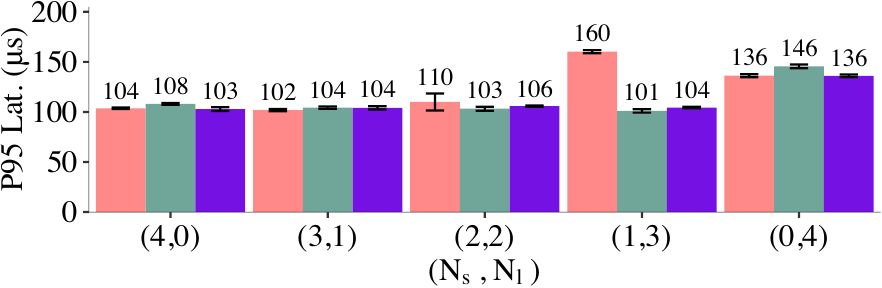} & 
\includegraphics[width=3in]{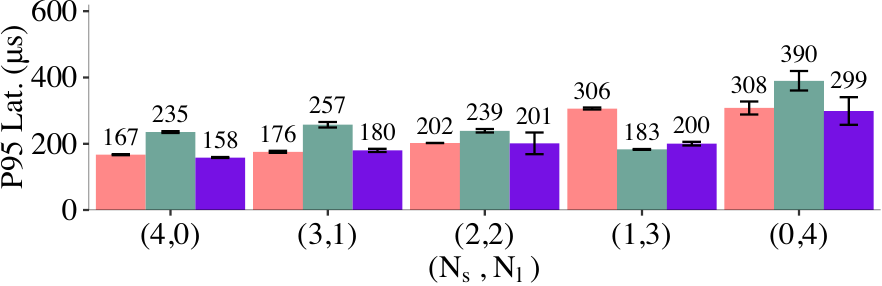} \\ 
{\small (e) 4\,KiB writes, p95 latency} &
{\small (f) 8\,KiB writes, p95 latency} 
\vspace{6pt}\\
\includegraphics[width=3in]{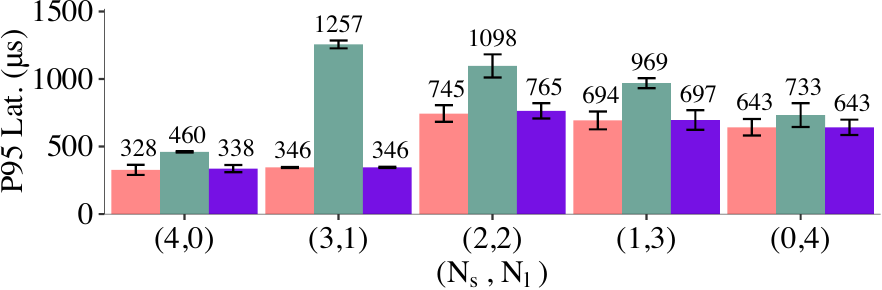} & 
\includegraphics[width=3in]{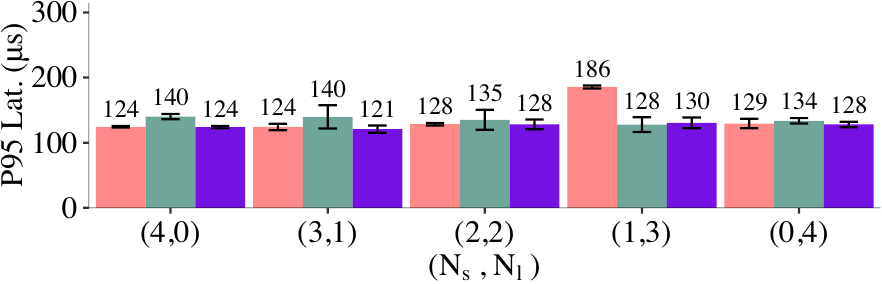} \\
{\small (g) 16\,KiB writes, p95 latency}&
{\small (h) Hybrid 4\,KiB and 16\,KiB writes, p95 latency} 
\vspace{6pt}\\
\includegraphics[width=3in]{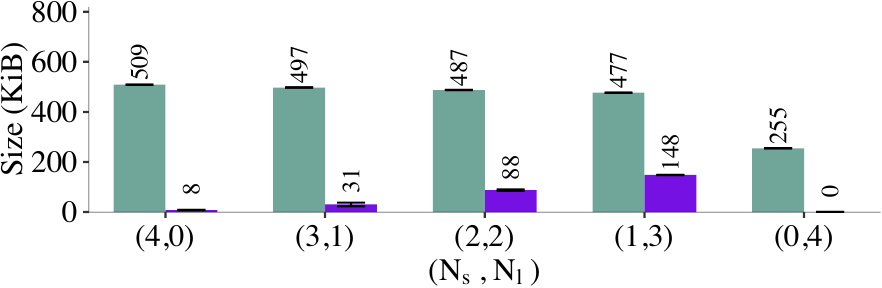} & 
\includegraphics[width=3in]{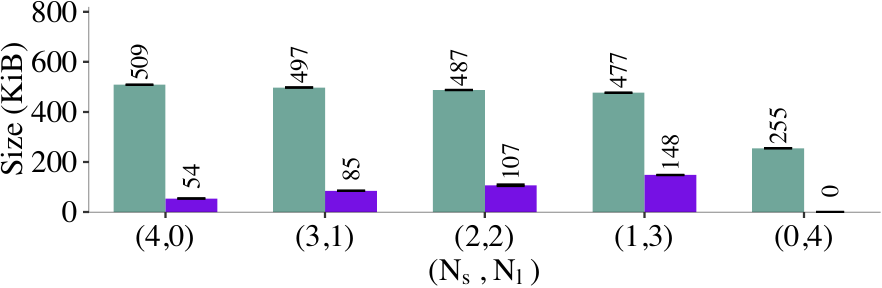} \\
\mbox{\small (i) 4\,KiB writes, memory size} &
\mbox{\small (j) 8\,KiB writes, memory size} 
\vspace{6pt}\\
\includegraphics[width=3in]{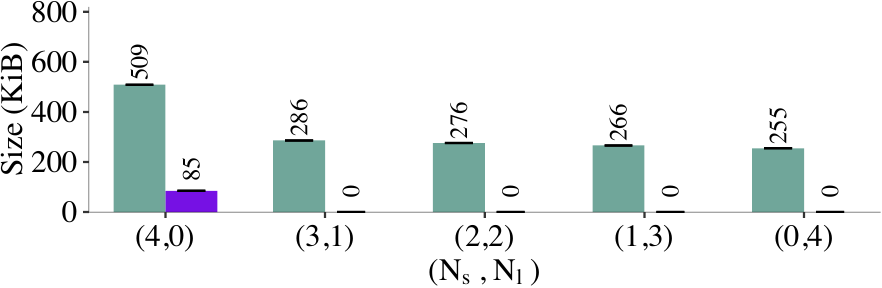} & 
\includegraphics[width=3in]{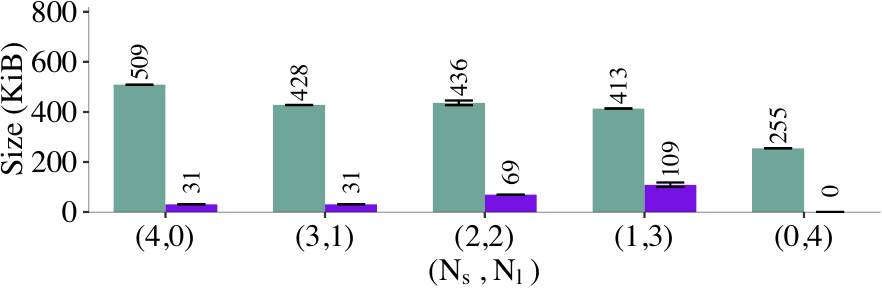} \\
\mbox{\small (k) 16\,KiB writes, memory size} &
\makecell[c]{\small(l) Hybrid 4\,KiB and 16\,KiB writes, memory size}
\end{tabular}
\vspace{-6pt}
\caption{Exp\#11 (Multiple open segments). We show the throughput
(figures~(a)-(d)), 95th-percentile latencies (figures~(e)-(h)), and
stripe management memory usage per GiB of written data (figures~(i)-(l))
under four open segments.}
\label{fig:write_4zones}
\end{figure}

Figures~\ref{fig:write_4zones}(a)-\ref{fig:write_4zones}(d) show the
throughput results.  ZoneWrite-Only and ZoneAppend-Only cannot always achieve
the highest throughput in all cases.  For example, ZoneAppend-Only has 65.7\%
higher throughput than ZoneWrite-Only for 4\,KiB writes and $(N_s, N_l)
= (1, 3)$ due to better intra-zone parallelism under small writes
(Figure~\ref{fig:write_4zones}(a)), but ZoneWrite-Only has 27.2\% higher
throughput than ZoneAppend-Only for 16\,KiB writes and $(N_s, N_l) = (1, 3)$ 
(Figure~\ref{fig:write_4zones}(c)).  On the other hand, \sysname achieves the
highest throughput, or similar throughput to the best scheme, in all cases
by carefully combining Zone Append and Zone Write. 

Note that while ZoneAppend-Only can issue 16\,KiB writes to multiple
large-chunk segments, it cannot achieve higher throughput as $N_l$ increases,
which may be inconsistent with the findings in Figure~\ref{fig:motivation}
(\S\ref{subsec:zns}). One possible reason is that the computational
overhead of the firmware for Zone Append becomes heavier and has higher
fluctuations when we issue writes via Zone Append to multiple drives, and
the performance is bottlenecked by the current slowest drive.  Note that
Zone Write also has a similar problem, but is less severe than Zone Append.
Such performance fluctuations are also observed in prior work
\cite{zhang18,jiang21}. 

Figures~\ref{fig:write_4zones}(e)-\ref{fig:write_4zones}(h) show the
95th-percentile latency results. Overall, \sysname maintains low
95th-percentile latencies in all cases.  For 4\,KiB writes
(Figure~\ref{fig:write_4zones}(e)) and hybrid 4\,KiB and 16\,KiB writes
(Figure~\ref{fig:write_4zones}(h)), all three schemes have similar latencies
except for $(N_s, N_l) = (1, 3)$, ZoneWrite-Only has a significantly high
latency since the 4\,KiB writes are issued to the only small-chunk segment and
ZoneWrite-Only has limited intra-zone parallelism. 

For 8\,KiB writes (Figure~\ref{fig:write_4zones}(f)), \sysname and
ZoneWrite-Only have similar 95th-percentile latencies, while \sysname has
15.9-32.7\% lower 95th-percentile latencies than ZoneAppend-Only for 
$(N_s, N_l)=$~(4,0), (3,1), and (2,2) by better exploiting inter-zone
parallelism through both Zone Append and Zone Write.  For $(N_s, N_l)=(1, 3)$,
\sysname has a slightly higher 95th-percentile latency than ZoneAppend-Only
(by 9.4\%) due to the overhead of group-based data layout. 

For 16\,KiB writes (Figure~\ref{fig:write_4zones}(g)), \sysname and
ZoneWrite-Only have consistently lower 95th-percentile latencies than
ZoneAppend-Only.  We observe that the 95th-percentile latency of
ZoneAppend-Only grows sharply by 173.3\% from $(N_s, N_l)=(4,0)$ to $(N_s,
N_l)=(3,1)$ since the 16\,KiB writes are issued to the only open large-chunk
segment for $N_l=1$ while they are issued to four open small-chunk segments
for $N_l=0$.  The 95th-percentile latency of issuing 16\,KiB writes to a single
large-chunk segment is much higher (see also Figure~\ref{fig:write}(c) in
Exp\#5). As $N_l$ further increases from one to four, the 95th-percentile
latency of ZoneAppend-Only drops by 58.3\% as there are more open large-chunk
segments and a higher degree of inter-zone parallelism.  We also observe that
the 95th-percentile latencies of \sysname and ZoneWrite-Only increase by
121.1\% and 115.3\%, respectively from $(N_s, N_l) = (3,1)$ to $(N_s,
N_l)=(2,2)$. The reason is that the Zone Write command on one open segment has
low on-SSD computational overhead as it only handles one outstanding write
request.  The overhead grows when it issues multiple outstanding write
requests to different open segments. Similar to ZoneAppend-Only, their
95th-percentile latencies drop as $N_l$ increases from two to four due to a
higher degree of inter-zone parallelism. 

Figures~\ref{fig:write_4zones}(i)-\ref{fig:write_4zones}(l) show the
stripe management memory size per GiB of written data.  We focus on
ZoneAppend-Only and \sysname as in Exp\#5.  \sysname maintains significantly
lower memory usage with its compact stripe table design than ZoneAppend-Only
in all cases. For example, for 4\,KiB writes
(Figure~\ref{fig:write_4zones}(i)), \sysname achieves 68.9-98.5\% lower memory
usage than ZoneAppend-Only for $(N_s,N_l) = (4,0)$, $(3,1)$, $(2,2)$, and
$(1,3)$.  Also, \sysname incurs no memory usage for the compact stripe table
for the cases of (i) $(N_s, N_l) = (0, 4)$ and (ii) $N_l > 0$ and there are
only large writes (i.e., 16\,KiB writes), since it only issues Zone Write to
large-chunk segments.  Thus, \sysname keeps low or no memory usage for stripe
management, while ZoneAppend-Only always incurs significant memory usage for
tracking the mappings between stripes and chunk locations.

We briefly examine the performance impact of separating small and
large writes to small-chunk and large-chunk segments, respectively in
\sysname.  Specifically, we consider the case of hybrid 4\,KiB and 16\,KiB
writes.  From Figure~\ref{fig:write_4zones}(d), when we change $(N_s,N_l)$
from $(4,0)$ to $(3,1)$, we move the 16\,KiB writes from small-chunk segments
to a large-chunk segment (recall that $(N_s, N_l) = (4,0)$ means that all
writes are issued to small-chunk segments), and the throughput of \sysname
increases from 3,181\,MiB/s to 3,459\,MiB/s.  Using a large $N_l$ can further
increase the throughput (e.g., the throughput reaches 3,583\,MiB/s for $(N_s,
N_l) = (0,4)$ when all writes are issued to large-chunk segments), but it
significantly increases the tail latency.  Here, we examine the
99th-percentile latencies (as the 95th-percentile latencies are similar from
Figure~\ref{fig:write_4zones}(h)). 
We observe that the 99th-percentile latency of \sysname increases from
168.8$\mu$s for $(N_s,N_l)=(4,0)$ to 171.2\,$\mu$s for $(N_s,N_l)=(3,1)$, and
further significantly increases to 342.2\,$\mu$s for $(N_s, N_l) = (0,4)$.
Note that the focus of this paper is to appropriately use Zone Write and Zone
Append for a given $(N_s, N_l)$.  Determining the best $(N_s, N_l)$
configuration is orthogonal to our work and left for future work.

\begin{figure}[!t]
\centering
\begin{tabular}{@{\ }cc}
\multicolumn{2}{c}{
\includegraphics[width=1.4in]{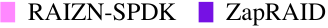}}\\
\includegraphics[width=3in]{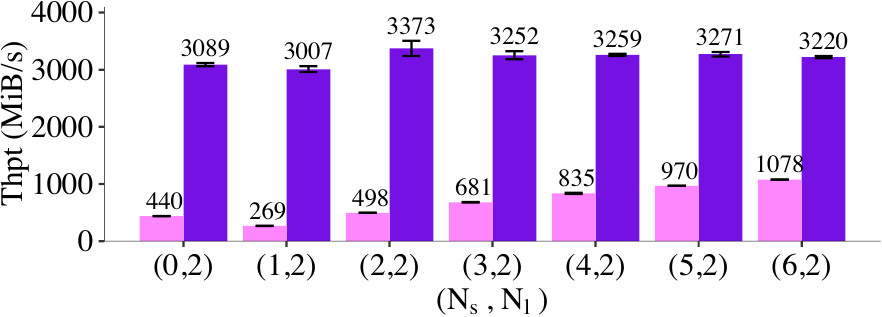} & 
\includegraphics[width=3in]{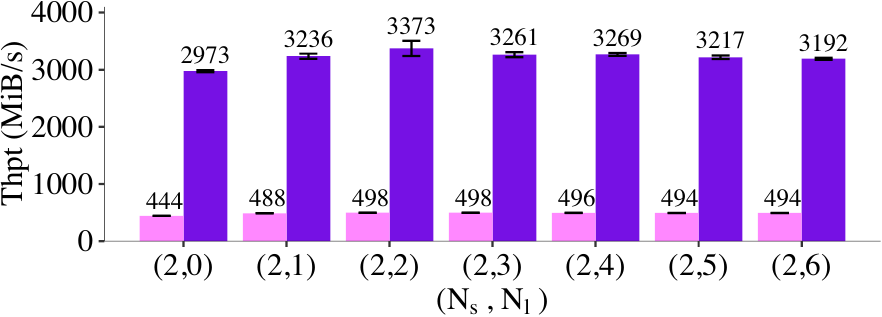} \\ 
{\small (a) $N_l=2$ and varying $N_s$, throughput} & 
{\small (b) $N_s=2$ and varying $N_l$, throughput} 
\vspace{6pt}\\
\includegraphics[width=3in]{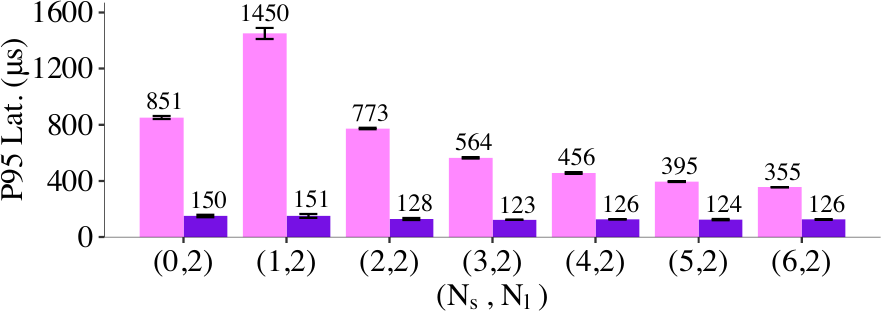} &
\includegraphics[width=3in]{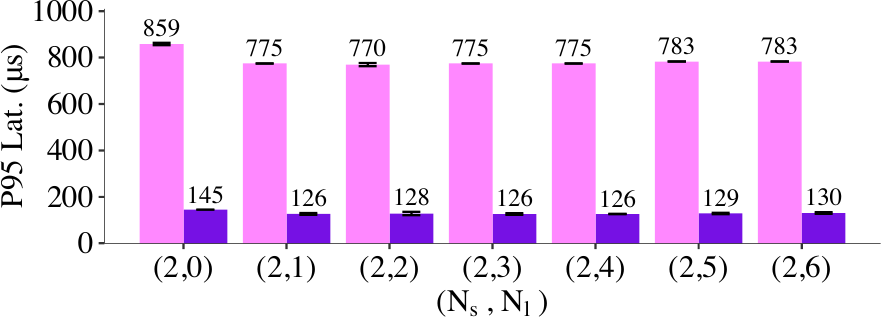} \\
{\small (c) $N_l=2$ and varying $N_s$, p95 latency} & 
{\small (d) $N_s=2$ and varying $N_l$, p95 latency} 
\end{tabular}
\vspace{-6pt}
\caption{Exp\#11 (Multiple open segments). We compare the throughput
and 95th-percentile latencies of RAIZN-SPDK and \sysname under hybrid 75\%
4\,KiB and 25\% 16\,KiB writes and different combinations of $(N_s, N_l)$.}
\label{fig:raizn_multiple}
\end{figure}

We also compare \sysname with RAIZN-SPDK under multiple open segments.
We focus on the workloads with hybrid 75\% 4\,KiB and 25\% 16\,KiB writes, and
examine different combinations of $(N_s, N_l)$.  We consider two settings: 
(i) we fix $N_l=2$ and vary $N_s$ from 0 to 6, and (ii) we fix $N_s=2$ and
vary $N_l$ from 0 to 6.

Figure~\ref{fig:raizn_multiple}(a) shows the throughput results of RAIZN-SPDK
and \sysname when we fix $N_l=2$ and vary $N_s$.  For RAIZN-SPDK, when $(N_s,
N_l)$ changes from $(0,2)$ to $(1,2)$, its throughput drops from 440.4\,MiB/s
to 268.6\,MiB/s.  The reason is that RAIZN-SPDK forwards all 4\,KiB writes to
two large-chunk segments for $(N_s,N_l)=(0,2)$, but for $(N_s,N_l)=(1,2)$, it
forwards all 4\,KiB writes to only one small-chunk segment and fails to
leverage inter-zone parallelism for these small writes.  When $(N_s,N_l)$
changes from $(1,2)$ to $(6,2)$, the throughput of RAIZN-SPDK increases to
1,077.6\,MiB/s with the increasing inter-zone parallelism across small-chunk
open segments.  For \sysname, with the help of intra-zone parallelism for
small writes, it maintains similar throughput for $(N_s,N_l)=(0,2)$ and 
$(N_s,N_l)=(1,2)$, even though the latter has only one small-chunk open
segment.

Figure~\ref{fig:raizn_multiple}(b) shows the throughput results of RAIZN-SPDK
and \sysname when we fix $N_s=2$ and vary $N_l$.  The throughput of RAIZN-SPDK
is 444.6\,MiB/s for $(N_s,N_l)=(2,0)$, but only increases to
488.4-498.0\,MiB/s when $(N_s,N_l)$ varies from $(2,1)$ to $(2,6)$.  The
reason is that most of the write requests (75\% in total) are 4\,KiB writes,
and the performance is bottlenecked by the limited number of small-chunk open
segments.  

Figures~\ref{fig:raizn_multiple}(c) and \ref{fig:raizn_multiple}(d) show the
95th-percentile latency results of RAIZN-SPDK and \sysname, and the
observations are consistent with those in the throughput results.  Overall,
when there are multiple open segments, \sysname still achieves significantly
higher throughput and lower latencies than RAIZN-SPDK, which incurs large
overhead of partial parity updates.

\begin{table}[!t]
\small
\centering
\renewcommand{\arraystretch}{1.15}
\setlength{\tabcolsep}{4pt}
\begin{tabular}{c|c|c|c|c|c||c|c|c|c}
\hline
 & {\bf ($N_s, N_l$)} & {\bf (0, 2)} & {\bf (1,2)} & {\bf (2,2)} & {\bf (6,2)} &
    {\bf (2,0)} & {\bf (2,1)} & {\bf (2,2)} & {\bf (2,6)} \\ 
\hline
\multirow{3}{*}{\sysname} & {\bf Wait phase} & 27.01 & 31.92 & 34.38 & 41.01 &
    39.10 & 35.89 & 34.38 & 38.17 \\
\cline{2-10} & {\bf Data phase} & 72.86 & 74.30 & 66.86 & 67.25 &
    76.31 & 66.23 & 66.86 & 70.27 \\
\cline{2-10} & {\bf Parity phase} & 5.44 & 3.69 & 3.68 & 3.52 &
    3.43 & 3.76 & 3.68 & 3.60 \\
\hline
\multirow{3}{*}{RAIZN-SPDK} & {\bf Wait phase} & 775.46 & 1282.24 & 678.87 & 292.05 & 
    861.28 & 694.47 & 678.87 & 682.73 \\
\cline{2-10} & {\bf Data phase} & 23.66 & 22.62 & 23.96 & 31.46 & 
    26.08 & 23.83 & 23.96 & 24.24 \\
\cline{2-10} & {\bf Parity phase} & 1.01 & 1.02 & 1.02 & 1.11 & 1.04 & 1.04 & 1.02 & 1.04 \\
\hline
\end{tabular}
\vspace{3pt}
\caption{Exp\#11 (Multiple open segments). Breakdown of latencies (in
microseconds) of \sysname and RAIZN-SPDK for
Figure~\ref{fig:raizn_multiple}.}
\label{tab:raizn_breakdown}
\end{table}

Table~\ref{tab:raizn_breakdown} further shows the latency breakdown
of RAIZN-SPDK and \sysname based on some of the settings of $(N_s, N_l)$ in
Figure~\ref{fig:raizn_multiple}, so as to explain the large overhead of partial
parity updates in RAIZN-SPDK.  For each write request, we record four
timestamps: (i) $t_1$, the time when the request is issued by fio, (ii) $t_2$,
the time when the request starts writing data chunks, (iii) $t_3$, the time
when the request finishes writing data chunks, and (iv) $t_4$, the time when
the request completes writing all parity chunks (including partial parity
updates for RAIZN-SPDK).  We divide each write request into three phases: 
(i) the {\em wait} phase (i.e., $t_2 - t_1$), (ii) the {\em data} phase (i.e.,
$t_3 - t_2$), and (iii) the {\em parity} phase (i.e., $t_4 - t_3$).  The table
shows that RAIZN-SPDK has a prolonged wait phase, as each of its write
requests must update the parity chunks and wait for the completion of the
partial parity updates of the {\em previous} write request for the same
segment, thereby limiting the concurrency of writes.  In contrast, \sysname
has a short wait phase as it can issue writes concurrently (recall that our
default queue depth is 64).  Note that \sysname has longer data and parity
phases than RAIZN-SPDK since its concurrent writes may overlap, so on average
each write request takes longer to complete.  Nevertheless, the overall write
throughput of \sysname is significantly higher than that of RAIZN-SPDK.

\begin{figure}[!t]
\centering
\begin{tabular}{@{\ }c@{\ }c@{\ }c}
\multicolumn{3}{c}{
\includegraphics[width=5in]{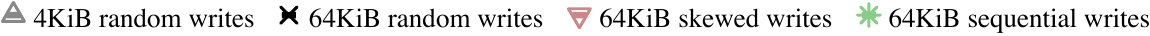}
}\\
\includegraphics[width=2in]{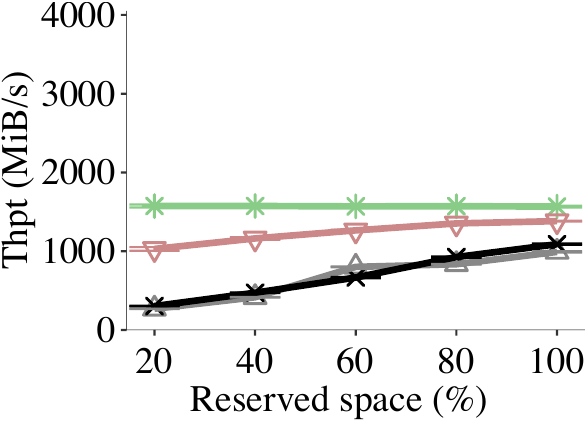} & 
\includegraphics[width=2in]{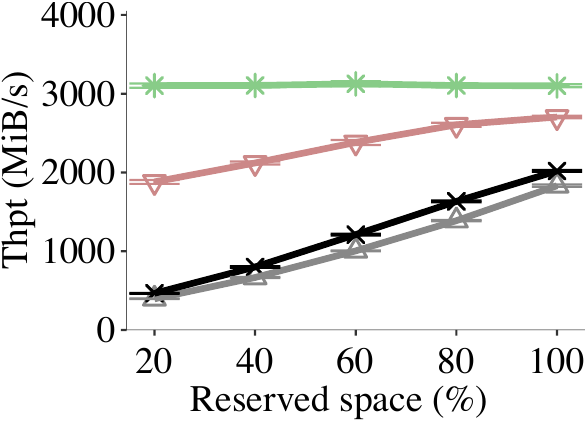} & 
\includegraphics[width=2in]{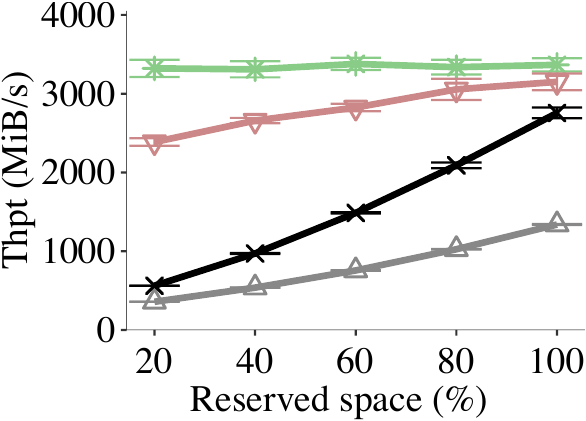} \\ 
{\small (a) 4\,KiB chunks} &
{\small (b) 16\,KiB chunks} &
{\small (c) Multiple open segments}
\end{tabular}
\vspace{-6pt}
\caption{Exp\#12 (Overhead of garbage collection).}
\label{fig:eval_gc}
\end{figure}

\para{Exp\#12 (Overhead of garbage collection in \sysname).} We evaluate
the impact of garbage collection on \sysname with different reserved space
sizes.  We issue writes to 200\,GiB of data, with a total of 1\,TiB of write
traffic.  We vary the physical space capacity of \sysname from 240\,GiB to
400\,GiB, meaning that the reserved space for garbage collection varies
between 20\% and 100\%.  We consider four write workloads, including 4\,KiB
(uniform) random writes, 64\,KiB (uniform) random writes, 64\,KiB skewed writes,
and 64\,KiB sequential writes.  We consider 64\,KiB writes to examine the
overhead of garbage collection caused by much larger writes.  For 64\,KiB
skewed writes, we consider a Zipfian distribution with a skewness factor of
0.99.  We consider three settings: (i) a single open segment with 4\,KiB
chunks, (ii) a single open segment with 16\,KiB chunks, and (iii) four open
segments with $(N_s, N_l) = (2,2)$, $C_s=$~8\,KiB, and $C_l=$~16\,KiB.

Figure~\ref{fig:eval_gc} shows the results. As expected, the write throughput
increases (i.e., the garbage collection overhead is less) with more reserved
space and with more skewed or sequential workloads. Since real-world workloads
are often skewed in practice (e.g., block storage \cite{yang16} and key-value
\cite{atikoglu12,cao20} workloads), we expect that the garbage collection
overhead is acceptable in \sysname.  Note that the write throughput under
multiple open segments (Figure~\ref{fig:eval_gc}(c)) is higher than under
single open segments (Figures~\ref{fig:eval_gc}(a) and \ref{fig:eval_gc}(b)),
meaning that the garbage collection overhead can be reduced via inter-zone
parallelism. 

\para{Exp\#13 (Overhead of L2P table offloading).} We evaluate the
overhead of offloading the L2P table entries to ZNS SSDs
(\S\ref{subsec:overview}). We consider the same four write workloads as in
Exp\#12 and focus on the setting of $(N_s, N_l)=(2, 2)$ with four open
segments.  We set the physical space capacity as 300\,GiB and issue writes to
200\,GiB of data (i.e., the overprovisioned space is 50\%). We vary
the maximum memory size allocated for the in-memory L2P table from 50\,MiB to
200\,MiB.  Recall that each L2P table entry has a size of four bytes for each
4\,KiB block (\S\ref{subsec:overview}).  Thus, the in-memory L2P table can store
the entries from 50\,GiB to 200\,GiB of data.  Note that when the in-memory L2P
table is allocated with 200\,MiB, it can keep all L2P table entries in memory
(i.e., no offloading). 

\begin{figure}[!t]
\centering
\includegraphics[width=4.6in]{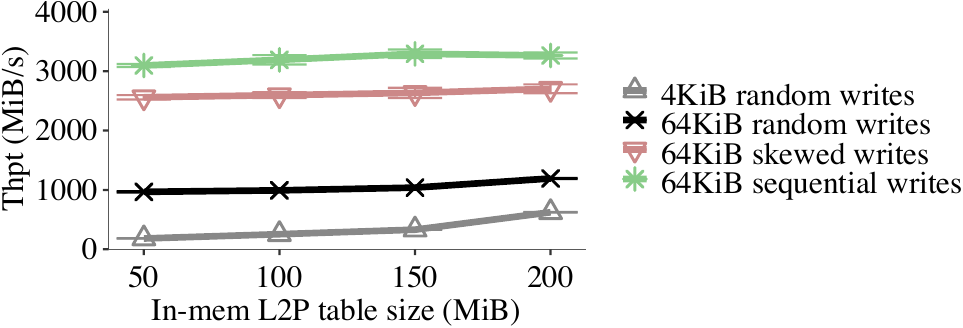}
\caption{Exp\#13 (Overhead of L2P table offloading).}
\label{fig:eval_l2p}
\end{figure}

Figure~\ref{fig:eval_l2p} shows the results. When the in-memory L2P table has
less than 200\,MiB of memory space, it needs to offload some entries onto the
SSD.  For 4\,KiB and 64\,KiB random writes, a 100\,MiB L2P table has lower
throughput than a 200\,MiB L2P table by 59.2\% and 16.8\%, respectively.  The
reason for such high throughput drops is that under uniform random writes, the
chance of not caching an L2P table entry is higher.  Nevertheless, larger
writes have lower degradations since \sysname groups the L2P table entries
into mapping blocks and there will be less I/O overhead of reading L2P table
entries from ZNS SSDs.  For 64\,KiB skewed writes, a 100\,MiB L2P table only has
slightly lower throughput than a 200\,MiB L2P table by 4.0\%, meaning that the
offloading overhead becomes insignificant in skewed workloads.  For 64\,KiB
sequential writes, a 100\,MiB L2P table only has slightly lower throughput than
a 200\,MiB L2P table by 3.6\%. 

\subsection{Summary of Findings}
\label{subsec:obsv}

\sysname is designed to integrate the strengths of
both Zone Append and Zone Write primitives into a RAID system instead of
always outperforming either in isolation.  It is particularly effective for
workloads with only small writes (e.g., 4\,KiB or 8\,KiB) or mixed workloads
with most small writes (e.g., 4\,KiB or 8\,KiB) and few large writes (e.g.,
16\,KiB).  This pattern is common in practice.  For example, in the Alibaba
Cloud traces (Exp\#4), 684 out of 1,000 volumes (68.4\%) contain more than
60\% of requests no larger than 4\,KiB and more than 25\% of requests at least
16\,KiB, with writes forming the majority of I/O. 

Compared to ZoneWrite-Only, \sysname improves the write throughput for small
writes in a single open segment, with a trade-off of introducing an extra
compact stripe table.  However, \sysname limits the size of the compact stripe
table with the group-based data layout and ensures that the query overhead
during degraded reads is minimal.  While ZoneWrite-Only shows lower p95
latency in large writes in a single open segment (e.g.,
Figures~\ref{fig:write}(c) and \ref{fig:write_4kreq}(c) in Exp\#5), \sysname
introduces hybrid data management to use Zone Write for large writes and
multiple open segments.  When multiple open segments are available, \sysname
shows no worse p95 write latency (Figure~\ref{fig:write_4zones} in Exp\#11).

Compared to ZoneAppend-Only, \sysname matches the write throughput in a single
open segment because \sysname uses the Zone Append primitive for a single open
segment. However, ZoneAppend-Only shows significantly poor degraded read
performance (Exp\#2) and higher memory usage for index management (Exp\#3) in
real-application experiments.  \sysname outperforms ZoneAppend-Only with high
degraded read performance (which is comparable to ZoneWrite-Only) and limited
memory usage for index management.

We discuss the implications of our findings in terms of performance,
memory usage, and functionalities of \sysname, compared with ZoneWrite-Only
and ZoneAppend-Only.

\para{Performance.} \sysname demonstrates superior write performance and
comparable normal read performance compared with ZoneWrite-Only and
ZoneAppend-Only.  Also, \sysname achieves significantly lower degraded read
latencies than ZoneAppend-Only. 
\begin{itemize}[leftmargin=*]
\item 
In terms of write performance, \sysname shows the highest write throughput
under single-segment settings, as shown in Exp\#1-Exp\#4, Exp\#5, and Exp\#8,
particularly with 4\,KiB chunks, in which \sysname improves
write throughput by up to 77.2\% (Exp\#5). Also, \sysname shows the highest
write throughput and lowest write tail latencies across various configurations
of $(N_s, N_l)$ under multiple-segment settings (Exp\#1-Exp\#4 and Exp\#11).
Furthermore, \sysname excels under highly concurrent write workloads (Exp\#1
and Exp\#10).
\item 
In terms of read performance, \sysname improves degraded read performance with
lightweight stripe management.  It maintains comparable degraded read
performance compared with ZoneWrite-Only (Exp\#6 and Exp\#7) and has
significantly higher degraded read performance than ZoneAppend-Only
(Exp\#2 and Exp\#7).
Note that \sysname is designed to maintain (rather than improve) normal read
performance, so its overall throughput improvements are low for read-intensive
applications (Exp\#1 and Exp\#4). 
\end{itemize}

\para{Memory usage.}  \sysname substantially reduces memory usage via its
compact stripe table design compared with ZoneAppend-Only. It further reduces
memory usage by offloading the L2P table to drives. 
\begin{itemize}[leftmargin=*]
\item 
For the compact stripe table, \sysname reduces the compact stripe table size
in two aspects. First, the group-based data layout (\S\ref{subsec:group})
significantly reduces the compact stripe table size for each zone by 66.7\%
(Exp\#3 and Exp\#5). Second, the hybrid data management
(\S\ref{subsec:hybrid}) leverages Zone Write for multiple-segment settings, so
as to eliminate the need for the compact stripe table in those segments 
(Exp\#3 and Exp\#11). 
\item 
For the L2P table, \sysname offloads the L2P table to drives to further reduce
memory usage (Exp\#13), with negligible overhead for highly skewed or
sequential workloads. 
\end{itemize}

\para{Functionalities.} \sysname supports multiple RAID schemes, crash
recovery, full-drive recovery, and garbage collection, and hence adapts
effectively to various scenarios.
\begin{itemize}[leftmargin=*]
\item 
\sysname maintains the highest write performance across various RAID schemes
compared with ZoneWrite-Only and ZoneAppend-Only (Exp\#8). 
\item 
\sysname maintains high performance in crash recovery and full-drive
recovery (Exp\#9).
\item 
\sysname supports garbage collection by recycling overwritten or discarded
user-written blocks. Its garbage collection overhead is small for highly
skewed or sequential write workloads (Exp\#12).
\end{itemize}

\section{Related Work}
\label{sec:related}

\noindent
{\bf SSD RAID.}  SSD RAID has been extensively studied in the literature.
Some studies follow traditional disk-based RAID (with in-place updates)
\cite{patterson88} and focus on reducing read tail latencies \cite{hao16},
improving parity update performance \cite{im11, kim13, chung14, li16,jiang21},
optimizing garbage collection \cite{kim11, kim12, wu18}, making I/O
performance predictable \cite{li21}, and enhancing the scalability of Linux
software RAID \cite{wang22,yi22}.  To improve scalability, some studies
organize RAID by distributing the controller function to multiple nodes
\cite{cao93} and further reducing the network traffic under disaggregated
storage \cite{shu23}. 
In this work, we focus on Log-RAID architectures, which have been extensively
studied for SSD RAID to improve both write performance and flash endurance. 
SOFA \cite{chiueh14} places the FTL in the RAID controller for efficient data
management.  Purity \cite{colgrove15} manages both indexing and data storage
under Log-RAID, and supports compression and deduplication.  SALSA
\cite{ioannou18} implements a general translation layer for SSDs and Shingled
Magnetic Recording (SMR) disks.  SWAN \cite{kim19} proposes spatial data
separation to reduce garbage collection interference.  Recent studies also
apply RAID on Key-Value SSDs (KVSSDs) \cite{pitchumani20,maheshwari20,qin21}
and open-channel SSDs \cite{ma22}.  \sysname targets RAID for ZNS SSDs and
particularly exploits Zone Append for high performance.

\para{Storage systems for ZNS SSDs.} Several studies explore new storage
system designs for ZNS SSDs.  Bjorling {\em et al.} \cite{bjorling21} adapt
F2FS \cite{lee15} and RocksDB \cite{rocksdb} for ZNS SSDs, and show that the
adapted systems have lower I/O amplification and higher performance than
conventional SSDs.  Other studies focus on the performance and management
aspects of ZNS SSDs. Examples include improving the performance of
log-structured merge-tree stores \cite{oneil96} for ZNS SSDs
\cite{jung22,lee22,lee23waltz}, optimizing host-level garbage collection
\cite{choi20,han21,seo23,byeon23}, proposing new ZNS interfaces for efficient
zone management \cite{maheshwari21,han21,min23}, enabling ZNS SSDs for swap
storage \cite{bergman22}, designing new I/O scheduling for improved intra-zone
parallelism \cite{bae22}, extending Zone Append for sub-block data appends
\cite{purandare22}, and ensuring crash consistency on F2FS backed by ZNS SSDs
\cite{lee23}.

While the above studies focus on a single ZNS SSD, RAIZN \cite{kim23} exposes
a ZNS SSD array as a single ZNS interface to applications, and focuses on
fault tolerance, correctness, and crash consistency. BIZA \cite{yi24}
presents a self-governing block-interface ZNS all-flash-array that hides the
ZNS interface from applications, while internally exploiting zone random write
area (ZRWA) and intra-/inter-zone parallelism to improve both endurance and
performance.  In contrast, \sysname focuses on exploiting Zone Append for high
performance. 

\para{Interfaces for offloading address management.}  While our work
focuses on the Zone Append interface, some studies propose new storage
interfaces that offload address management from applications to storage
devices as in Zone Append.  Examples include range writes \cite{anand08} and 
Direct File System \cite{josephson10} for file system management, nameless
writes \cite{zhang12} and software-enabled flash \cite{sef} for flash-based
SSDs, as well as ZEA \cite{manzanares16} for host-managed shingled magnetic
recording drives.  We conjecture that \sysname, which builds on Zone Append,
can also build on the above interfaces to improve RAID performance, and we
pose this issue as future work. 

\section{Conclusion}

\sysname is a high-performance Log-RAID system for ZNS SSDs to support
scalable and reliable storage, with the design goals of high performance,
lightweight stripe management, and reliability in mind.  We propose the
following key design elements: (i) a group-based data layout under Zone Append
to effectively exploit intra-zone parallelism while mitigating the stripe
management overhead; (ii) hybrid data management to carefully combine Zone
Append and Zone Write to achieve intra-zone and inter-zone parallelism; (iii)  
the crash consistency mechanisms to enforce high reliability.  Our prototype
evaluation on real ZNS SSDs and FEMU emulation shows that \sysname achieves high
write throughput and maintains efficiency in degraded reads, crash recovery,
and full-drive recovery.  
\end{sloppypar}

\bibliographystyle{plain}
\bibliography{reference}

\end{document}